\documentstyle[aps,pre,psfig]{revtex}

\def\be{\begin{equation}}
\def\beq{\begin{equation}}
\def\bea{\begin{eqnarray}}
\def\ee{\end{equation}}
\def\eeq{\end{equation}}
\def\eea{\end{eqnarray}}
\def\to{\rightarrow}

\def\ra{\rangle}
\def\la{\langle}
\def\r{\right}
\def\l{\left}
\def\a{\alpha}
\def\b{\beta}
\def\g{\gamma}

\def\max{{\rm max}}
\def\mig{{\rm mig}}
\def\rep{{\rm rep}}
\def\ini{{\rm ini}}

\def\bds{\begin{displaystyle}}
\def\eds{\end{displaystyle}}

\begin{document}
\tightenlines

\title{Diversity patterns from ecological models at dynamical equilibrium}
\author{U. Bastolla$^1$, M. L\"assig$^2$, S.C. Manrubia$^1$, and A. 
Valleriani$^1$}
\address{$^1$Max Planck Institute of Colloids and Interfaces, 
Theory Division, D-14424 Potsdam (Germany) \\ 
$^2$ Institute for Theoretical Physics, University of Cologne, Z\"ulpicher 
Strasse 77, D-50937 Cologne (Germany)}
\date{\today}

\maketitle

\begin{abstract}
We study a dynamic model of ecosystems where immigration plays an essential
role both in assembling the species community and in mantaining its
biodiversity. This framework is particularly relevant for insular ecosystems.
Population dynamics is represented either as an individual based model or as 
a set of deterministic equations for population abundances. Local extinctions 
and immigrations balance in a statistically stationary state where 
biodiversity fluctuates around a constant mean value. At stationarity, 
biodiversity increases as a power law of the immigration rate.
Our model yields almost power-law
species-area relationships, with a range of effective exponents in agreement
with that observed for biodiversity of whole archipelagos.
We also observe broad distributions for species abundances and species
lifetimes and a small number of trophic levels, limited by the immigration
rate. These results are rather robust with respect to change of description 
level, as well as change of population dynamic equations, from prey dependent 
to ratio dependent.
\end{abstract}

\section{Introduction}

One of the most beautiful instances of statistical laws in biology is the 
species-area law, which relates the area $A$ of a habitat and the 
number of different species $S$ coexisting there. In its qualitative form, 
this law has already been stated by Alexander von Humboldt in the 19th-century:
Larger areas harbor more species than smaller ones (see Rosenzweig, 
1999). The most commonly used  quantitative  relation between 
number of species and area has the form of a power-law,
\be
\label{SAz}
S \propto A^z\;.
\ee
The exponent $z$  depends on the geographical characteristics
of the region under consideration and on the taxon considered
(see, however,  He \& Legendre, 1996 for a different view).

Despite the large number of studies on biodiversity patterns,
their relationship to underlying population dynamic models
has hardly been explored until now.  This is the scope of the present paper.
We obtain  scaling laws for biodiversity as a function of 
external control parameters like the immigration rate and the total 
amount of abiotic resources.
Relating these external parameters to the area of our model island,
we obtain a species area relationship that compares favorably
simplifito field data. 

In this approach, biodiversity is  established
and mantained by an immigration flux, much like in the phenomenological
theory of island biogeography (MacArthur \& Wilson, 1963).
At long times,  the system always evolves to a
statistically {\em stationary state} where immigrations and local extinctions
balance.
It is then characterized by a constant
turnover of species, but global quantities do not change on the average,
and ecological variables like the number of
species, the number of trophic levels, the number of links per species or
the species abundances attain a time-independent distribution. 
This state can be called a dynamical equilibrium (MacArthur \& Wilson, 1963).
We emphasize, however, that a biosystem in this state is {\em strongly driven}:
its biodiversity depends on a nonzero flux of immigrations.
It cannot be described by a given fixed point of population dynamics. The
static description of ecosystems, very common in classical population
dynamics, seems thus inadequate to model biodiversity.

A large number of field studies 
support the functional relation given by Eq.\ (\ref{SAz}), with exponents
which range from $0.13$ to $0.18$ for nested areas in the mainland, in the 
interval $0.25-0.33$ for groups of
nearby islands, go up to $0.5-0.8$ for archipelagoes, and range in 
$0.17-0.72$ for habitat islands (Rosenzweig, 1995, Begon {\it et al.}, 1998).

Searching for an explanation of insular biodiversity patterns,
MacArthur and Wilson (MacArthur \& Wilson, 1963) proposed in the sixties
an equilibrium  theory of island biogeography.
According to this theory,
the number of species on islands is the result of a dynamical 
balance between the arrival of new species (immigration) and local 
extinction. Many field studies and statistical analysis of data have
been carried out to define the applicability limits of
the theory. They range from island defaunation 
experiments (Simberloff  \& Wilson, 1969) and subsequent analysis 
(Simberloff, 1969, Heatwole \& Levins, 1972) to the study of insular
biodiversity patterns (Gilpin \& Diamond, 1976 )
and of fluctuations of the number of species (Gilpin \& Diamond, 1980, 1981;
Manne {\em et al.}, 1998).
On the other hand, the lack of explanatory power of MacArthur and
Wilson's theory has been criticized
(Williamson, 1989; Whittaker, 1992), and corrections hav been proposed
(Simberloff, 1969). The major shortcoming in our view is that the approach 
is not explicitly founded on ecological dynamics.

Many authors have investigated the species-area relationship without 
relying on the equilibrium theory of island biogeography.
A classical model was proposed by Preston (Preston, 1962), 
who derived a
power-law  relationship from the assumption that the abundance of
species is characterized by a log-normal distribution. 
This assumption however has been
questioned, since field study support broader distributions.
Recently, Harte and collaborators obtained a power-law relationship between
species diversity and area from the hypothesis that the spatial distribution
of individuals is self similar with respect to the operation of cutting a
small area from a larger one (Harte {\it et al.}, 1999). Thus
their model applies to continental nested subareas. An analytical relation
between the scale-invariant spatial distribution and the species 
distribution was subsequently derived (Banavar {\it et al.}, 1999). Still, the
hypothesis of a self-similar distribution of abundances is a strong
assumption, even if it seems supported by some field data. 

Several recent models combine immigration or speciation and ecological
dynamics, trying
to overcome the major shortcoming of the previous approaches.
Wissel modelled an ecosystem of similar species (Wissel, 1992). He 
combined the effects of environmental and demographic stochasticity
together with interspecies competition thus obtaining a power law species-area
relationship.
Durrett and Levin proposed a model where speciation is coupled to
contact dynamics to mimic ecological processes, obtaining nearly power
law species area relationships (Durrett \& Levin, 1996).
Caldarelli {\it et al.} (1998) and Drossel {\it et al.} (2000)
coupled population dynamics to speciation and immigration processes
to simulate changes in biodiversity over evolutionary times.
Loreau and Moquet modelled immigrations of plant species from a large
pool to an island (Loreau \& Mouquet, 1999). They 
included explicit competition for space in the framework of the equilibrium
theory.
A recent model for species turnover has reproduced power law distributions 
for species abundances and species lifetimes as observed in field studies, 
providing moreover a mechanism for switching
from power-law to log-normal distributions as parameters are changed
(Sol\'e {\it et al.}, 2000). The combination of a diffusion mechanism 
coupled to spatial noise (and leading to local extinctions) generates a 
power law relationship between number of species and area with exponent
$z \simeq 0.25$, even if interaction among species is not explicitly 
represented (Pelletier, 1999).

Another class of models which combine immigrations and
population dynamics is that of species assembly
(Post \& Pimm, 1983; Drake, 1990; Case, 1990; Morton \& Law, 1997;
Happel \& Stadler, 1998; Schreiber \& Gutierrez, 1998).
In these models, a community is constructed through local immigrations
from a regional species pool. After every immigration, the new community
is tested for persistence ({\it i.e.}, the property that no species gets
extinct even in the limit of infinite time).
Imposing the condition of persistence bounds these models to the
limit of very rare immigrations,
while in our model the immigration rate plays the role of the
essential control parameter.
\vspace{.2cm}

The classical view of ecosystems represents them as static ensembles
of species at a stable fixed point of population dynamics.
In this framework, a key result is constituted by May's 
theorem (May, 1972). He showed that a large ecological system formed by 
$S$ species
with a total number of $C$ random connections with average value zero and 
dispersion $\sigma$ will have, with probability one, no
stable fixed point if the variance of the interactions verifies $\sigma^2 >
C/S$. This result, which has been slightly corrected in more recent years
(Cohen \& Newman, 1985), sets an upper limit to the amount of complexity
allowed by the stability condition, and breaks the traditional view of
complexity as a force increasing (static) stability.

The representation of ecosystems as static entities was already challenged
by MacArthur and Wilson's theory. According to them
ecosystems are ever changing ensembles of species, evolving
in a stationary state where the average number of species
remains constant in time.
In the same spirit, we take a dynamical view of ecological processes.
We model immigrations to an insular ecosystem as a flux of species coming
from a continent. Starting from an empty island, new species arrive
at random and build the model food webs. This is a possible way of assembling 
the ecological community (see also Drake, 1990). Population 
dynamics leads ultimately to species extinctions, but the ecosystem stays 
in general far from any fixed point of the dynamics.

After a transient period, a statistically stationary state where extinctions 
balance new immigrations is reached. As will be seen,
this stationary state is rather far from static equilibrium: In fact,
if the immigration rate is switched off, we observe that several species go
rapidly extinct and the ecosystem reaches, more and more slowly as more
species go extinct, a static fixed point with a very small number
of species.
Although we used different models to describe immigrations and population
dynamics, we observed that the main statistical features of the equilibrium
state are the same for different descriptions. This makes us confident that 
our results are robust and fairly independent of the details of the
modellization.

The relationship between the time scale of the
external driving (the immigration rate) and that of the internal ecological
dynamics assumes in this context a key importance.
If the typical time scale for immigrations is very large compared
to the time scale for equilibration of the ecological dynamics, then
a new fixed point is reached after every new arrival. In this case,
immigrations act only as proposal of new species, but do not shape the
ecosystem, whose properties will be independent of the immigration rate. If, 
on the other hand, the immigration of new species is very fast compared to 
the ecological dynamics, the ecosystem will be determined only by 
immigrations and will resemble a random assemblage of species. At 
intermediate time scales non trivial equilibria emerge and both 
immigrations and ecological dynamics play an important role.
In this regime, the average number of species in the stationary
state, $S$, increases as a power law of the immigration rate $I$,
\be
S\approx I^a\:, \label{S-I}
\ee
with values for the exponent typically
in a narrow range, $0.4<a\leq 1$, depending on the system parameters. 

We shall relate Eq.\ (\ref{S-I}) to the species-area 
relationship in Sec. \ref{area}, where we assume that the immigration
rate is proportional to the linear size of the island
(MacArthur \& Wilson, 1963). The corresponding exponent $z$ depends on
parameter values, but is typically comprised in a narrow range,
from $z=0.52$ to $0.56$. These values are very close to those observed for
biodiversity in whole archipelagos, when one single source of immigrants
is considered. Since our model does not consider spatial structure, it is
implicitely assumed that our ``island" is either isolated or corresponds to
a whole archipelago. In this case, our results show a remarkable agreement 
with experimental data. 
It is thus tempting to speculate that the effect of area
on biodiversity is largely due to the immigration rate, and that the
exponent $z$ observed for islands of the same archipelago is smaller
because also interchanges of species have to be considered.
We shall argue that an approximately power law species-area relationship
could be a generic feature arising from the ecological dynamics and the
existence of a statistical stationary state, therefore independent of the
details of the system.

The immigration rate influences also other properties of the food
webs: The number of trophic levels increases with $I$,
consistently with the observation that food chain length is positively
correlated to the size of the ecosystem (Schoener, 1989) and with
a recent simulation (Spencer, 1997).
Also the number of links per species and the total biomass change with
the immigration rate.
The distribution of species abundances has a power-law shape, with an
exponent close to $-1$ and slowly decreasing with the
immigration rate. The first result is in agreement with field observations
(Pielou, 1969), considerations based on the theory of multiplicative processes
(Kerner, 1957; Sornette, 1998; Biham {\it et al.}, 1998)
and results from the simulations of a similar model
(Sol\'e {\it et al.}, 2000), and seems to be rather general.
The distribution of species lifetimes has also, in an intermediate range,
almost power law shape,
with an exponent close to $-2$ slowly decreasing with the
immigration rate, as it should be expected. Also in this case this is in
agreement with field
observations (Keitt \& Marquet, 1996; Keitt \& Stanley, 1998)
and with the results of the model by (Sol\'e {\it et al.}, 2000).

The paper is organized as follows.
In Sec. \ref{IBM} we introduce an Individual Based Model of ecological
dynamics based on stochastic dynamics. In Sec. \ref{L-V} we present
a formulation of ecological dynamics based on continuous deterministic
models. Since we are interested in the
comparison between these two description levels, the
results will be presented together in Sec. \ref{Results}. 
We conclude with an overall discussion.

\section{An individual based model}
\label{IBM}

Recently, population ecology has started to use individual based 
(or individual oriented) models 
(IBM) as a complementary tool in the study of ecological dynamics 
({\L}omnicki, 1999; Grimm {\it et al.}, 1999). One of the main interests 
of such approach is
that it allows the explicit modelization of individual characteristics, 
like the age of the individuals in a population 
(influencing the time of breeding or the moment at which they die), or the
energy that they store and require to move and survive (Bascompte {\it et 
al.}, 1997). Most
IBM studies refer to concrete problems where a few species 
of known characteristics interact to produce a well defined behaviour or
pattern, which the IBM should recover or predict 
(Fahse {\it et al.}, 1998; Spencer, 1997). 
Another interest of IBM is in what has been termed {\it virtual
ecology}: The comparison between real data and simulated data obtained
from a system where realistic restrictions have been considered might allow
the design of better protocols for recruitment and observation
(Berger {\it et al.}, 1999; Hall \& Halle, 1999).

Simulations of very large systems with many individuals
and/or many species have not been undertaken until recently because of
computational limitations. Thus the IBM approach was restricted to few species
in relatively small lattices representing real space, with one to few 
individuals per lattice site. More ambitious problems, like the 
relation between theoretical results for deterministic continuous models and
their IBM counterpart, were addressed only recently
(Keitt, 1997). Some authors derived time-continuous models from the more
basic description of the flow of energy between constituents (Svirezhev, 1997)
or among individuals (Wilson, 1998; Sol\'e {\it et al.}, 1999). This is 
indeed a very relevant point. One
would expect that the coarse-grained higher-level description represented by
deterministic models captures the essential features of lower-level
individual-based models.
This is in fact the phylosophy behind our approach: In the IBM, as 
well as in the higher-level models to be introduced in the forthcoming 
sections, we study the predictions of the model from a statistical point of 
view, ignoring details that will necessarily be different in different 
models.

\subsection{Ecological dynamics}

Consider a large area $A_{\rm cont}$ on which a maximum number $M_h$ of basal 
species coexist and up to $M_a$ animal species compete for resources. 
The ecological interactions in this community will be defined through a 
matrix with entries 
$C(m_i,m_j)$. Depending on the values of $C(m_i,m_j)$ and $C(m_j,m_i)$
we will determine the trophic relationship between individual $i$ of species 
$m_i$ and individual $j$ of species $m_j$, as we shall explain later.
We have considered two possible algorithms to
determine the non-zero elements of the interaction matrix. Our first 
election corresponds to the cascade model (Cohen {\it et al.}, 1990), which
returns a network with topological properties comparable to those of 
real ecosystems.
In this case, the distinction between basal and animal species automatically
arises from the ecological relationships given by the interaction matrix. 
We define $M=M_h+M_a$ to be the total number of species in the system. 
If the number $m_i = 1, \dots M$ specifies a peching order for 
feeding, the algorithm works as follows: Any species $m_i$ can feed only on 
species $m_j$ which is lower in the order, that is, $m_j < m_i$. This avoids
the formation of loops. A link to any of the potential prey species is 
established with probability $\ell/M$. If the value of $\ell$ is fixed
(according to real observations) to be around four, this model returns the 
correct proportions of basal, intermediate, and top species, a maximum 
number of levels typically around ten, and a distribution of the 
number of predators per prey which agrees with field observations 
(Cohen {\it et al.}, 1990). 

A second possibility for the interaction matrix consists in randomly assigning
$\ell$ preys to each of the $M_a$ animal species. This would correspond to 
a disordered situation where no processes have acted in order
to select the topology of the ecological network. In this case, and only for 
implementation purposes, basal species occupy positions $m_i=1, \dots
M_h$, and animal species occupy $m_i= M_h+1, \dots, (M_h+M_a)$. 
The interaction matrix has the form  

$$ \left( \begin{array}{c|c} 0 \;\;  &  0 \\ 
\hline C(m_a,m_h) & C(m_a,m_a) \end{array}
\right) $$
where $m_h$ and $m_a$ indicate basal species and animals,
respectively.
The statistical properties of the system do not depend on the form chosen 
for the interaction matrix. As
we will see, the relevant quantities take the same form in the 
cascade model case (CM) and in the random matrix case (RM).
For both algorithms, the values of the matrix elements are 
randomly chosen from a uniform distribution in $[0,E]$, where $E\in [20-200]$
(see Table I).
The value of the matrix
coefficients is proportional to the energy gained by individual $i$ when 
feeding on individual $j$ and represents a sort of assimilation efficiency
(see below).

In determining the matrix $C$, 
we have essentially defined a structured ecosystem in a very large
area with many species.
This is what we consider to be the continent,
which will be the source of immigration of propagules to an island of area 
$N_h$. This last quantity 
can be understood as the maximum number of patches covered by grass, for
instance, and acts as a limiting value (together with the basal growth, to be
defined) for the number of animals that will inhabit the island.

As time proceeds, and once we properly define the immigration mechanism, we 
will have a number $n_h(t)$ of individuals in basal
species present on the island and a number $n_a(t)$ of individuals belonging to
animal species. The total number of individuals in a wide sense (say patches
of grass plus animals) is
$n(t)=n_h(t)+n_a(t)$. Each individual is characterized by an energy
$e(i)$. Individuals reproduce provided their value of $e(i)$ is large enough.
Basal species increase their energy at a constant rate.
Animals dissipate energy as time elapses,
and increase the value of $e(i)$ through predation, which happens
stochastically. At each time-step the following rules are implemented:

\begin{enumerate}
\item {\bf Pair formation.} At each time step, we randomly form $n(t)/2$ 
pairs of individuals, independently of their specific affiliation.
If $n(t)$ is odd, one individual remains without partner.
This rule can be understood as a mean field picture of a space-explicit
approach. Different possibilities are i) $h-h$ pair: the two
grass patches are not consumed by animals and keep their energy,
ii) $a-h$ pair: if the matrix
element $C(m_a,m_h)$ is positive (meaning that the individual $a$ 
feeds on $h$), predation is possible, iii) $a-a$ pair, allowing predation 
between different animal species depending on the matrix coefficients.

\item {\bf Predation and Feeding.} Either of the individuals in each 
pair $(i,j)$ can feed on its partner, according to the 
ecological relations
defined in the matrix $C(m_i,m_j)$. Predation happens when
$C(m_i,m_j) \ne 0$ and $C(m_j,m_i) = 0$. In this case,

$$ \begin{array}{lll} e(i) & \to & e(i) + C(m_i,m_j) \times {e(j) \over
E_\rep} \\
e(j) & \to & 0  \; , \end{array} $$
where $E_\rep$ is an energy scale related to reproduction, that will be
defined below.
The energy received is proportional to the matrix element, but also to
the energy stored in the predated individual. In this sense, to eat a
new born is not equivalent to eating an adult close to its reproductive
energy (which fixes the maximum energy). Furthermore, an individual
with a total energy $e(i) > E+E_{rep}$ cannot further increase the value of
$e(i)$. In addition, if the value of the fraction $e(j)/E_{rep}$ is larger
than unity the rule is modified as $e(i) \to e(i) + C(m_i,m_j)$.

If both $C(m_i,m_j)$ and
$C(m_j,m_i)$ are non-zero (or both zero), no interaction takes place.

\item {\bf Basal growth.} Every individual belonging to a basal species
increases its energy at each time step by a net amount $b$,

$$ e(i) \to e(i) + b \; \; \; {\rm if} \; \; \; m_i \in [1,M_h] \; .$$

\item {\bf Dissipation.} At each time step, and for each of the alive
animals, $e(i) \rightarrow e(i) - d$, where $d$ defines the
dissipation rate. It takes the same value for all species in our model.

\item {\bf Reproduction.} If $e(i) \ge E_{rep}$, the individual $i$ is
allowed to reproduce. In the case of basal species, the new individual
is introduced into the system provided there is place, {\it i.e.} if
$n_h(t) < N_h$. The individual which reproduces loses an amount
of energy $\delta$,

$$ \begin{array}{lll} e(i) & \to & e(i) - \delta \\
e(k) & \to & \delta \; , \end{array}
$$
where $e(k)$ is the energy of the new born.

\item {\bf Death.} An individual can die for three different reasons:
If its energy reaches zero, if it is eaten by a predator, and
with a fixed probability $p_d$ per time step. 
\end{enumerate}
Table I resumes the parameters of the model and the approximate 
range of 
values used in our simulations. We will present results for some representative
cases. No qualitative differences were observed for comparable sets of
parameters.

\subsection{Modelling immigrations}

The initial
quenched matrix $C(m_i,m_j)$ can be thought of as the pool of species in the
continent, where a very large area (with its resources) allows
the coexistence of all possible species, in our case $M=M_h+M_a$. 
An island has a finite area $N_h$ and harbors only a subset $M_{\rm isl}$ of 
$M$. 

The immigration flux $I$ can take 
values from the set $\{ \dots \; 1/4, \; 1/3, \; 1/2, \; 1, \; 2, \; 3, \;
\dots\}$ only. If $I=q \ge 1$, then $q$ new individuals randomly chosen from 
any of the possible $M$ species in the pool arrive to the island at each 
time step. If $I = 1/q < 1$, one individual is introduced every $q$ time steps.
Other situations, which imply a less smooth flux are excluded in the 
following.

Our simulations show that the average number of species coexisting on
the island depends very strongly on the vertical 
transmission of resources, as is well known to happen in real ecosystems 
(Rosenzweig, 1995). High dissipation relatively to basal growth
($d$ close to $b$) turns into few species on the
island. For $d \ll b$ (a factor of 2 or 3 may suffice) the average
number of species coexisting when $I$ is large enough approaches the
maximum number $M$. 

With the addition of a constant flux of species from the continent to the
island, the system is poised to 
a state of dynamical equilibrium, where the number of species
that disappear due to the ecological interactions or to demographic 
stochasticity is balanced by the new incoming species. 
The immigration rate 
might produce a rescue effect for species with few individuals, close to 
extinction, and at the same time includes in a natural way one form of
environmental stochasticity. 

Thus, the incoming flux of individuals from the continent, the immigration
rate, becomes our main variable. By changing its intensity, we
can calculate the average number of species $S$ present in the 
statistically stable regime on an area $A \equiv N_h$. 
Moreover,
assuming a  relation between the immigration rate
$I$ and the area $A$, we shall derive the species-area law resulting from the
ecological dynamics of the IBM and compare it with field measurements.

\section{Deterministic continuous model}\label{L-V}

In this section we present the deterministic continuous models of population
dynamics adopted in our simulations. All individuals in each species are
grouped together and we represent them through a single dynamic variable,
the density of biomass (or abundance) of species $i$ at time $t$, $N_i(t)$. 

\subsection{Ecological dynamics}

The density of biomass of the species $i=1,\ldots, S$ evolve through a system of
differential equations,

\begin{equation}
{dN_i\over dt}=-\alpha_i N_i(t)-\beta_i \l(N_i(t)\r)^2 + \sum_j 
\gamma_{ij}N_j(t) N_i(t)\; ,
\label{equ}
\end{equation}
determining the growth rate of the biomasses as a function of the
abundances of all species in the ecosystem.

Species with biomass less than a predefined threshold value $N_c$ go extinct
and are eliminated from the system. This mimics the effect of demographic
stochasticity and the fact that species are made of discrete entities.

The term $-\alpha_i$ stands for the dissipation of energy following from
the biological activity of the members of species $i$ (movement, extraction
of nutrients, basal metabolism), as well as the death rate of individuals, and
corresponds to the quantities $d$ and $p_d$ of the IBM.
The term $-\beta_i N_i(t)$ is known as self-damping. 
It expresses a negative feedback of $N_i$ on its own growth rate, which
has been shown in some circumstances to be necessary in order to stabilize
the model.
The terms $\gamma_{ij}N_j$ represent the biomass transfered per unit time
from species $j$ to species $i$ if the sign is positive, and from species
$i$ to species $j$ if it is negative, thus modelling prey-predator
interactions. They are the counterpart of the matrix $C(m_i,m_j)$ in the IBM.

Energy flows into the system through the coupling of basal species to
external resources, which are formally represented as an additional 
``species'' $N_0$ whose equation will be specified below.
Terms of the form $\gamma_{i0} N_0$ are thus equivalent to the parameter $b$
in the IBM.
However modelled, external resources introduce in the system an energy scale
$R$ which limits its total biomass.

The quantity $g_{ij}\equiv \g_{ij} N_j$, equal to the energy transfer
from prey species $j$ to predator species $i$
per unit of predator, is called the predator's functional response
to prey $j$.
We studied two different variants of the continuous model, with different
functional responses and different equations for the resources.

\begin{itemize}

\item {\bf Model A}.
Generalized Lotka-Volterra
equations with constant $\g_{ij}$'s randomly drawn from a uniform distribution,
$\g_{ij} \in [0, \g_{\rm max}]$.

In order to represent competition among basal species,
we introduce a fictitious dynamics for the resources $N_0(t)$,
modelling them through an equation of the same kind as (\ref{equ}),

\be\label{Res2}
{1\over N_0}{dN_0\over dt}=\g_0R +\sum_j\g_{0j} N_j\; ,
\ee
where the constants $\gamma_{0j}$ have all negative signs and
we assume that at least one basal species is present.

The predator's functional response $g_{ij}=\g_{ij} N_j$
is proportional to prey biomass and belongs to the more general
category of prey dependent functional responses.

There is to observe that in Lotka-Volterra equations, the quantity
$\a/\g_\max$ introduces an energy scale in the ecosystem, aside to the
other energy scales $R$, $N_c$ and $\a/\b$. One can then expect that
different regimes are present for different relative values of these
energy scales, and this is indeed what we observe. Not all these regimes are
biologically meaningful. For instance, there is a regime, corresponding
to small values of the dimensionless parameter $u=\g_\max N_c/\a$, where
dissipation dominates and
only basal species can survive in the long run. We shall describe shortly
in the Appendix this garden regime.
More details on the regimes of
Lotka-Volterra equations will be given in a forthcoming work.

\item {\bf Model B}. Ratio dependent functional response.

Another possibility is that the energy scale of $1/\g_{ij}$
is determined by the biomasses $N_i$ and $N_j$. This choice is known
as {\em ratio dependent} predator response (Arditi \& Ginzburg, 1989),
since the functional response $\g_{ij}N_j$ depends on the ratio between the
prey biomass and the predator biomass.
In this case the $\g_{ij}$'s are not anymore constant, but they are inversely
proportional to some linear combination of $N_i$ and $N_j$.
The simplest possibility is that the prey $j$ has a unique predator $i$,
in which case the functional response is given by

\be g_{ij}(N_i,N_j)\equiv \gamma_{ij}(N_i,N_j)N_j=\l(a+b N_i/N_j\r)^{-1}\: .
\label{denom1}\ee

In the case of several predators for the prey $j$, different generalizations
of Eq.\ (\ref{denom1}) have been proposed (Arditi \& Michalski, 1995;
Schreiber \& Gutierrez, 1997; Drossel {\it et al.}, 2000).
We adopt our own generalization, which reads

\begin{equation} 
\gamma_{ij}= {\alpha'_{ij} c_{ij}\over b_j N_j+
\sum_{k\in P(j)}c_{kj}N_k}\: ,
\label{denom}
\end{equation}
where $i$ is the predator and $j$ the prey.
As in model (A), we assume $\g_{ij}=-\g_{ji}$.
Here $c_{ij}$ and $b_j$ are dimensionless coefficients,
$\alpha'_{ij}$ expresses the rate at which a single individual of species
$i$, in absence of competition, consumes a corresponding quantity of
biomass from species $j$, and $P(j)$ indicates the set of predators of
species $j$.

The above equations, unlike model (A), explicitly represent
the competition among predators of the same prey.
It is then possible to model external resouces as a constant flux of
energy available to basal species,

\be N_0(t)\equiv R \, . \ee

\end{itemize}

Ratio dependent and prey dependent functional responses have been supported
and criticized in several papers (see Abrams \& Ginzburg, 2000 for a recent 
review). We do not want to enter such a debate here.
Any functional response
is just a crude representation of a much more complicated situation, in
which spatial distributions of individuals, foraging strategies and mating
behaviour are involved. 
The point raised in this paper is that, although model (A) and model (B)
may have different scaling properties,
the scaling behaviour of biodiversity is robust with respect to changes
in the functional response, in an appropriate range of parameters. Indeed,
we observe that model (B) gives results qualitatively similar 
to those of model (A) in a range of $u=\g_\max N_c/\a$ where $S$ 
scales as $\log (R/N_c)$.

\subsection{Immigration and ecological parameters}

At time $t=0$ no species is present on the island. New species
arrive one after another, at fixed intervals of time,
$T_\mig$. Between successive arrivals, population dynamics
equations are integrated and species may go extinct.

For every new species, the ecological parameters are chosen at random
and kept fixed until the species goes extinct.
This means that new species are not related to species already
on the island, that is, the continental pool is considered infinite
respect to the number of species on the island.

New species have no predators on the island, and a number of preys
$\ell$ randomly extracted between one and $\ell_\max$ (in most
simulations we used either $\ell_\max=4$ or $\ell_\max=8$). 
The $\ell$ preys are extracted with uniform probability among the $S(t)$
existing species, regarding the external resources $N_0$
as a normal prey. This operation defines the ecological network.

For every link, the interaction strengths $\g_{ij}$ are extracted from
an uniform distribution in $[0,\, \g_\max]$ in the case of model (A).
In the case of model (B), the parameters $c_{ij}$ are extracted uniformly
in $[0,\, c_\max]$ and the remaining parameters are fixed at 
$\a'_{ij}\equiv 1$, $b_j\equiv 1$. In both cases, $i$ is the predator
and $j$ is the prey, and we then make the assumption that the
interaction strengths are antisymmetric: $\g_{ji}=-\g_{ij}$.
We also studied the case of reduced efficiency, $\g_{ij}=-\eta \g_{ji}$,
with $i$ predator, $j$ prey and $0.5<\eta <1$, without observing any
qualitative difference.
In case of model (A), the parameter $\g_0$, proportional to the growth
rate of the resources $N_0$, is set to $\g_0=\g_\max$.
As a simplification, the dissipation parameters are the same for
all species $\alpha_i=\alpha$, $\beta_i=\beta$.

Colonizing species arrive with very small populations. This assures that
they are rapidly eliminated if they are not fit for the island ecosystem.
We used initial values $N_\ini= N_c$ and $4N_c$. Reducing
the initial size reduces spurious effects due to species with very short
permanence in the system, and has only a very small influence on the 
statistical patterns described later.

\subsection{Discussion of the modelling choices}

The choice that new species have preys but not predators on the island
has to be justified.
This rule is aimed at forbidding the formation of ecological loops.
Moreover, we want that newcomers have considerable chances of surviving,
otherwise the rate of arrival of species with non vanishing permanence time
would have large fluctuations, increasing the fluctuations of all ecological
variables.
With this rule, the resulting ecological networks
resemble very much those obtained with the application of the cascade
algorithm, used in the IBM model.

Our simulations considered several representations
of the ecological dynamics at the individual and at the population level.
Although our consistent results let us believe that the models 
capture generic properties of ecological networks, there are a
number of alternative (and equally plausible) dynamical rules or
generalizations of our rules that would be interesting to consider. We
shortly discuss some of them.
 This is of course a strong
simplification. On the other hand, extracting at random for each species
parameters $\a_i$ and $\b_i$ would lead the average value of these
parameters to decrease towards zero, due to the advantage conferred by
small dissipation
rates and self-damping. In order to avoid this effect,
we believe that it is necessary to model a trade-off between dissipation,
$\a_i$, and predation efficiency $\g_{ij}$, in such a way that the latter
is an increasing function of the former. Another critical point is that
all pairs of coefficients $\g_{ij}$ and $\g_{ji}$ have, in our model, opposite
sign, so that symbiotic relationships are not represented.

Concerning the network structure, it is certainly a big simplification
to build links independently one of each other. For a more
realistic model one needs some measure of the distance among species in some
multidimensional space. It would then be possible, for instance, to extract
at random the first prey of a new species, and then to extract the
remaining preys with a probability depending on the distance from the
first prey. Another simplification, in the framework of the continuous
model, is that a new species has only preys and not predators. This was
imposed in order to avoid the formation of ecological loops, and is very
similar to the procedure adopted in the cascade model for network formation.
It is thus conforting that ecological networks constructed
both using the cascade model and random matrices containing loops (in the 
IBM framework) returned the same qualitative behaviour.

For very long times, coevolution of species would become relevant. This might 
be taking into account by making appropriate modifications
in the ecological network or in the interaction
parameters. Discarding coevolution is justified if the time scale
of the simulation is much shorter than the time after which at least one
species in the ecosystem mutates. Nevertheless, the latter time scale
is expected to decrease as the number of species in the ecosystem, $S(t)$,
increases, until a point where it is not anymore possible to neglect
coevolution. Such a situation is worth considering and will be the subject 
of future work.

\section{Statistical features of the stationary state}\label{Results}

For every set of observations,
we shall present both results obtained with the continuous model and
results obtained with the IBM, when available. In fact the two descriptions
produce the same qualitative behavior, even if simulations are much faster
for the continuous model, so that it is possible to simulate
larger systems and to obtain better statistics.

When we extract at random an ecological network with a high number of
species (up to 1000) and a low number of links per species (for instance four),
the ecological dynamics leads to the extinction of most of the species,
until only very few ones are represented in the system.
This result does not
seem to depend on the way in which the links have been extracted (either using
random matrices or through the cascade model), on the
parameter values, or on the kind of ecological dynamics represented
(individual based or continuous, with constant $\g_{ij}$ or with ratio
dependent response functions).

\begin{figure}
\centerline{
\psfig{file=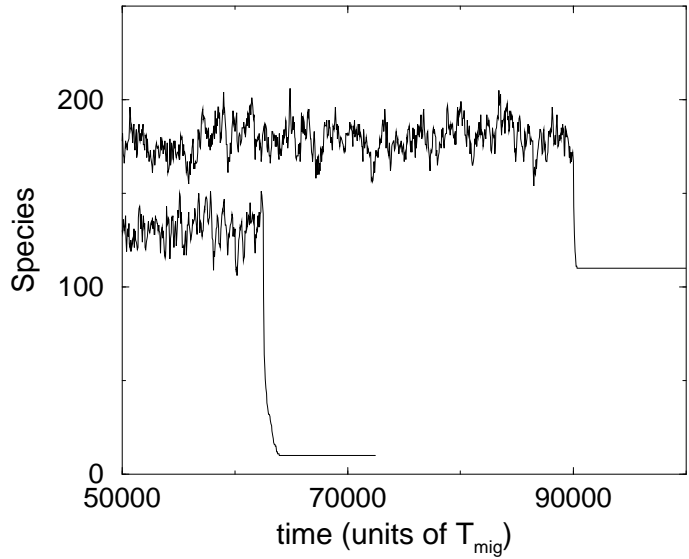,width=9cm}}
\caption{Number of species as a function of time in two typical realizations
of the continuous model.
The upper curve was obtained with Eq.\ (\ref{equ}) 
using the following parameters: $N'_c=10^3$, $\b'=10^4$, $\g'_\max =5000$,
$\ell_\max=8$, $T'_\mig=7\cdot 10^{-4}$. The lower curve was obtained with 
model (B) and parameters
$N'_c=10^4$, $\b'=10$, $c'_\max =20$,
$\ell_\max=8$, $T'_\mig=7 \cdot 10^{-3}$.
In both cases, only the final part of the evolution is shown. After a very
steep and fast transient ($t< 1000$), the system reaches a dynamical
equilibrium state. The steep decrease of biodiversity corresponds to
stopping the immigration rate. This last part and the initial transient part
of the curves were not used to measure stationary quantities.}
\label{fig:stat}
\end{figure}

In presence of a constant flux of immigrant species, the ecosystem,
initially empty, increases very fast in diversity until it reaches a
number of species which remains on the average stationary in time, although
characterized by large fluctuations. This process is illustrated in
Fig.\ \ref{fig:stat}. The system differs significantly from a static
network: In fact, if we stop immigrations we notice an abrupt
decrease in the number of species until a fixed point of much lower diversity 
is reached (see Fig.\ \ref{fig:stat}).

Choosing as time unity the quantity $1/\a$ and as biomass unity
the external resources $R$,
the dynamical equations can be written as a function of four
dimensionless parameters. For model (A) they are:
\be
\gamma_\max'=\gamma_\max R/\alpha  \;\;, \;\;\; N_c'=R/N_c \;\;, \;\;\;
\b'=\beta R/\alpha \;\;, \;\;\; T_\mig'=\alpha T_\mig\, .
\label{ParaA}
\ee
For model (B) the first parameter is substituted by
\be
c_\max'=c_\max/\alpha \, .
\label{ParaB}
\ee
Together with $\ell_\max$, the maximal number of preys for a
colonizing species, these parameters determine the model ecosystem.

\begin{figure}
\centerline{
\psfig{file=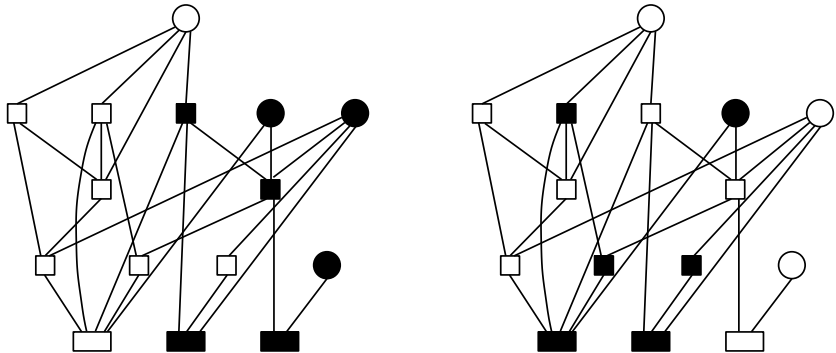,width=11cm,angle=0}}
\vspace{.5cm}
\caption{Two different ecological subnetworks for the IBM model
observed on the island for
the same network in the continent. The specific composition might 
be different at different time moments as a result of the turnover of 
species and
of the alternating dominance of different basal species. Here we show the
result of a simulation of a 
small ``continent'' with a total number of 15 species. Within an interval of
3000 time steps, the two configurations represented by filled symbols
(for species actually present on the island), were observed.
Circles stand for top species, squares for intermediate, and rectangles for 
basal ones.}
\label{fig:IBM1}
\end{figure}

We observe turnover of species in the system, 
and even a complete change in the species composition of the island in the 
course of time.
In the IBM model, where we use a fixed continental pool of species,
the presence of different basal species determines the intermediate and top 
species allowed by the subnetwork on the island. Due to stochastic effects, 
we observe an alternation between different basal species (often after a long
time interval), and consequently a complete renewal of the island ecosystem
(see Fig.\ \ref{fig:IBM1}).
This picture agrees qualitatively with experiments on island repopulation 
(Simberloff \& Wilson, 1969; Heatwole \& Levins, 1972;
Simberloff, 1969), where, after
defaunation, a different specific composition was obtained.

\subsection{Species diversity and relation among time-scales}

When a stationary state is achieved, we observe that, for some range
of parameters, the
average number of species $S$ increases as a power law of the immigration
rate $I=1/T_\mig$,

\be S\approx S_0+c I^a \; , \label{S-I0}
\ee
with a constant $S_0$ usually very small.

We define a new exponent $b$, that we call competition exponent, as

\be b={1\over a}-1. \label{comp}
\ee
In the stationary state, the average extinction rate per species is
$e=1/(T_\mig S)$ and increases with the number of species at
equilibrium as $S^b$ (for $S\gg S_0$ when only the immigration rate
increases), whence
the name of competition exponent:

\be
\label{comp2}
e \propto S^b. 
\ee

Since the exponent $b$ is larger than zero, the larger the number of
species, the smaller the time scale for the extinction of a single species.
The fact that the extinction rate increases with the number
of species has been postulated in the theory of island biogeography.
However, we find this result not as a phenomenological law, but as a generic
feature of the dynamics of random ecological networks.
We shall first present results from the IBM and then compare them
with those from the continuous models, Eq.\ (\ref{equ}).

\begin{figure}
\centerline{
\psfig{file=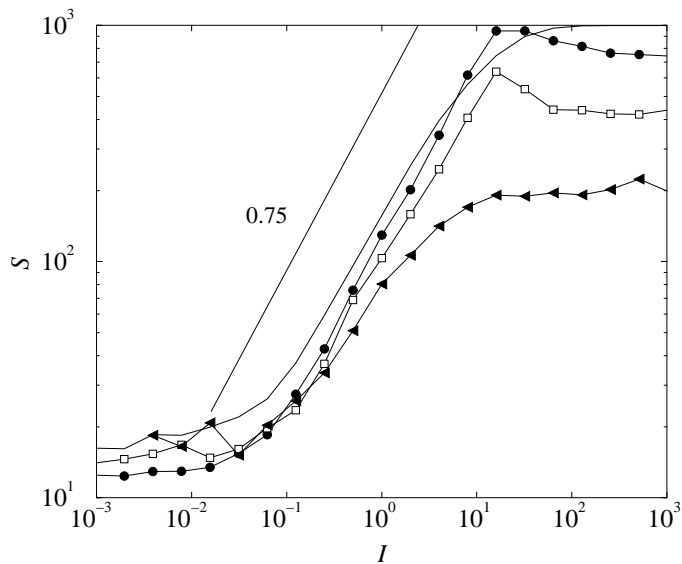,width=9cm,angle=270}}
\caption{Average number of species at equilibrium in the IBM model
as a function of the immigration rate $I$. Three different
regimes are observed as $I$ is increased. The
parameters for each of the curves are: 
$N_h=10^4$, $M_h=100$, $M_a=900$,
$E=100$, $E_{rep}=200$, $\ell=3$, $b=5$, $d=2$, $\delta=20$, 
$p_d=0.002$ (solid triangles); empty squares correspond to the same 
parameters with $d=1$, and solid circles to the same situation but  
$E=150$, $b=10$, and $d=1$. This three curves have been obtained from random
matrices (RM). The solid line corresponds to a simulation of the CM with 
parameters as in the last case.}
\label{fig:IBM2}
\end{figure}

The results of the IBM model are summarized in Fig.\ \ref{fig:IBM2}. Four
curves of the average number of species $S$ as a function of the
immigration rate $I$ are plotted.  Curves qualitatively similar to ours were 
obtained in other simple models for island colonization (Rosenzweig, 1995; 
Loreau \& Mouquet, 1999), but the functional relationship between $I$ and
$S$ was not investigated.

All curves in Fig.\ \ref{fig:IBM2} show two plateaus corresponding to a
i) low diversity regime (small immigration rate),
and ii) disordered species composition (large immigration rate),
which are linked by a 
transition region with power-law shape described by Eq.\ (\ref{S-I0}). 
The effective exponent obtained from a power-law fit is $a \simeq 0.75$.
This intermediate regime
is the most interesting one, since here both the immigration rate and the
ecological organization play a relevant role in setting the average number of
species at equilibrium. In the lower plateau, the dynamics is dominated by the
internal ecological processes, while in the higher plateau the fast
arrival of new individuals controls the species composition on the island. 
This latter situation is analogous to the one observed in the defaunation
experiment reported in (Simberloff \& Wilson,
1969; Simberloff, 1969). 

\begin{figure}
\centerline{
\psfig{file=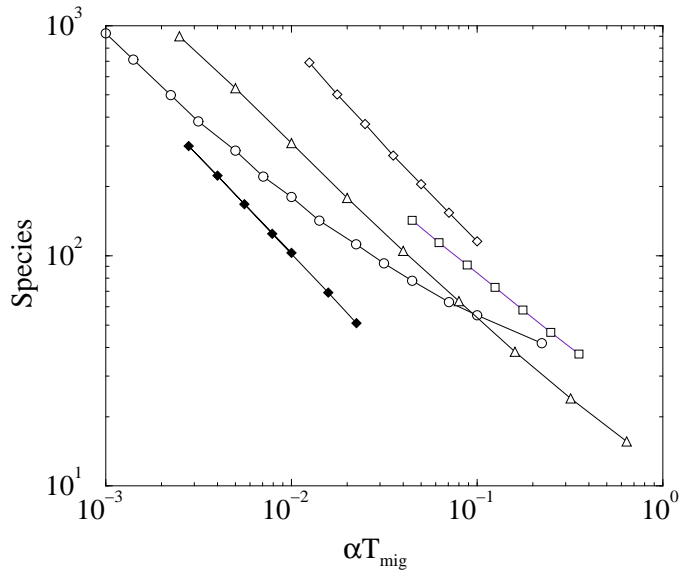,width=9cm}}
\vspace{.5cm}
\caption{Sample of plots of the stationary number of species as a function of
the inverse of the immigration rate, $T_\mig$, in units of $1/\a$.
Filled symbols refer to simulations of model (A) Eq.\ (\ref{equ}), empty
symbols are for model (B). Parameters are as follows.
Filled diamonds: $N'_c=10^3$, $\b' =10^3$, $\g'_\max =5\cdot 10^3$
(in this case the horizontal axis is $10\a T_\mig$).
Empty circles: $N'_c=10^5$, $\b'=10^2$, $c'_\max=20$.
Empty diamonds: $N'_c=10^4$, $\a=0$, $c_\max/\b R=20$
(in this case the horizontal axis is $T_\mig/\b R$).
Empty squares: $N'_c=3.2\cdot 10^6$, $\b'=32$, $c'_\max=2$.
Empty triangles: $N'_c=10^5$, $\b'=2$, $c'_\max=8$.
$\ell_\max$ is either four or eight. In all cases, $N_\ini=4N_c$.}
\label{fig:alpha}
\end{figure}

The results for the continuous model are completely similar. The regime between
the two plateaus is represented in Fig.\ \ref{fig:alpha}. 
Notice that in this case we do not have a
fixed continental pool, or, in other words, the parameter $M$ has to be
interpreted as infinite.
Some of the curves show a curvature in log-log plots, which can be
eliminated introducing $S_0$ as additional fit parameter and plotting
($S-S_0$) versus $I$. The effect of $S_0$ is thus to reduce the effective
exponent $a$ as the immigration flux is reduced.

The observed exponents range from $a=0.42$ to $a=1$.
Interestingly, the case with $a=1$, corresponding to a competition exponent
$b=0$ (not represented in Fig.\ \ref{fig:alpha}), refers to a case where
basal species were not in competition, since we used constant parameters
$\g_{ij}$ and constant resources $N_0(t)\equiv R$. In all other cases
the exponent $b$ was positive.

We now discuss shortly the behavior of biodiversity with the
parameters of the ecological equations. Keeping fixed the other
dimensionless parameters given in Eqs.\ (\ref{ParaA}) and (\ref{ParaB}),
biodiversity increases logarithmically with the resources $R/N_c$:

\be S\approx A_1+A_2\log\l(R/N_c\r)\, \label{S-log}.
\ee
This result holds for the IBM and for the continuous models, but for
model (A) it is only valid in some range of parameter values.
In fact, model (A) at fixed $\g'_{max}$ is found, for large $R/N_c$, in
the garden regime, where only basal species survive, and the scaling
behavior is different there (see Appendix).
The exponent $a$, defined in Eq.\ (\ref{S-I0}), changes only very weakly
with $R/N_c$.

Biodiversity also increases with the maximal transfer
rate, either $c'_\max$ for the case of Eq.\ (\ref{denom}) or $\g'_\max$
for model (A), when all other parameters in Eq.\ (\ref{ParaA}) are kept 
constant. In the first case, the number of species tends to zero as 
$c'_\max$ approaches unity (for $N_\ini=N_c$). In the second case, at 
small $\g'_\max/N'_c$ we reach
the ``garden regime'' (see Appendix)
where the number of species is almost independent of $T_\mig$.
Both limits can be interpreted as corresponding to $a=0$, thus $b=\infty$.
The exponent $a$ then increases slowly with $\g_\max$.

The effect of the parameter $\b'$, when other parameters in
Eqs.\ (\ref{ParaA}) and (\ref{ParaB}) are fixed, depends on the immigration
rate. While biodiversity increases for increasing
$\b'$ at small immigration, the opposite happens if immigration is large. 
Thus, $I-S$
curves relative to different values of $\b'$ should cross at some point.
This behavior reflects on the fact that the exponent $a$ decreases with
increasing $\b'$,  while $S_0$ increases.
As in the case of $\g'_\max$, the decrease of $a$ can be explained by the fact
that, at larger $\b'$, the probability that a colonizer has a positive
growth rate becomes smaller. The positive effect on $S_0$, on the other
hand, is due to the fact that the larger $\b'$, the more likely  that
two predators feeding on the same prey species can coexist.

\begin{figure}
\centerline{
\psfig{file=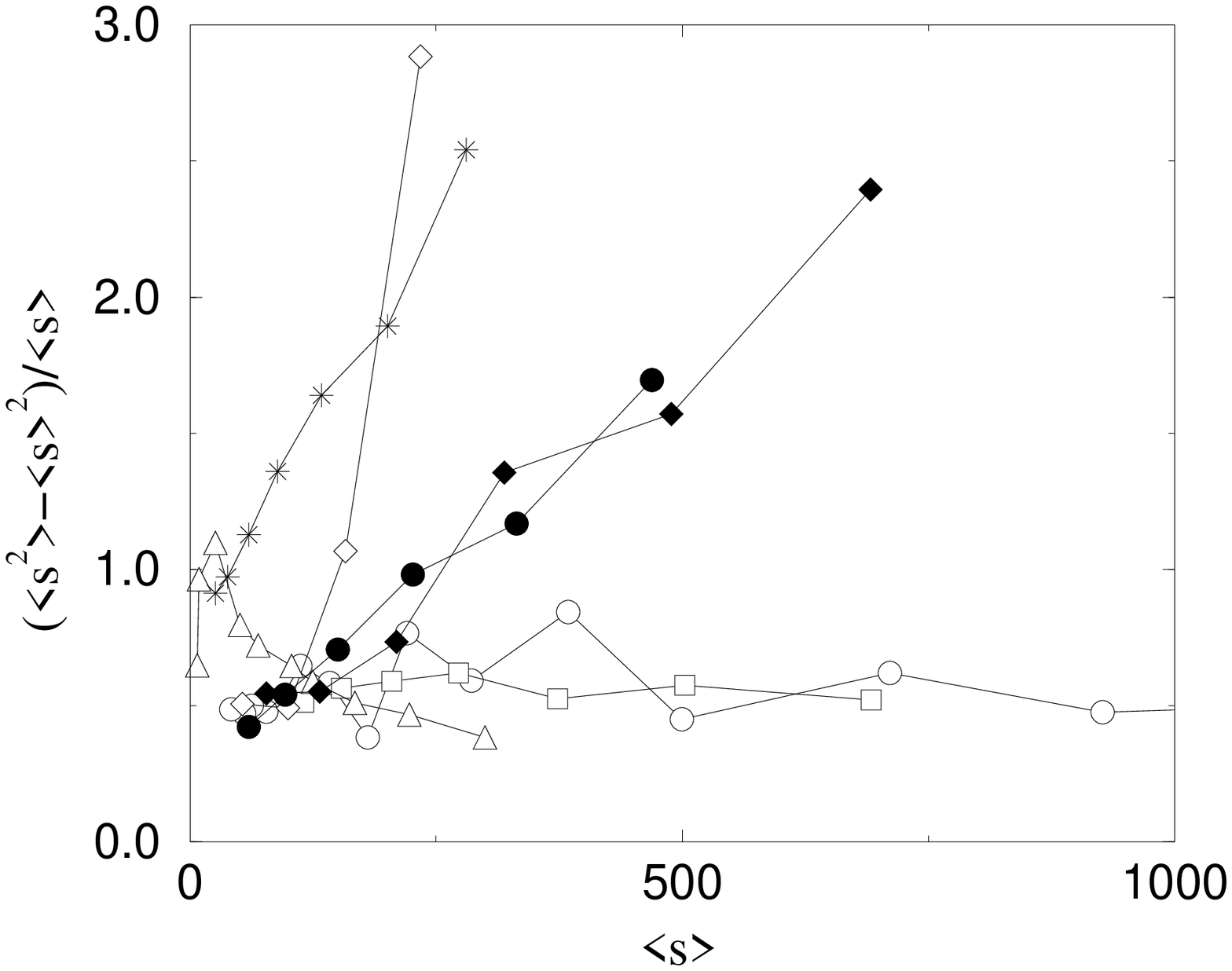,width=8cm}
\psfig{file=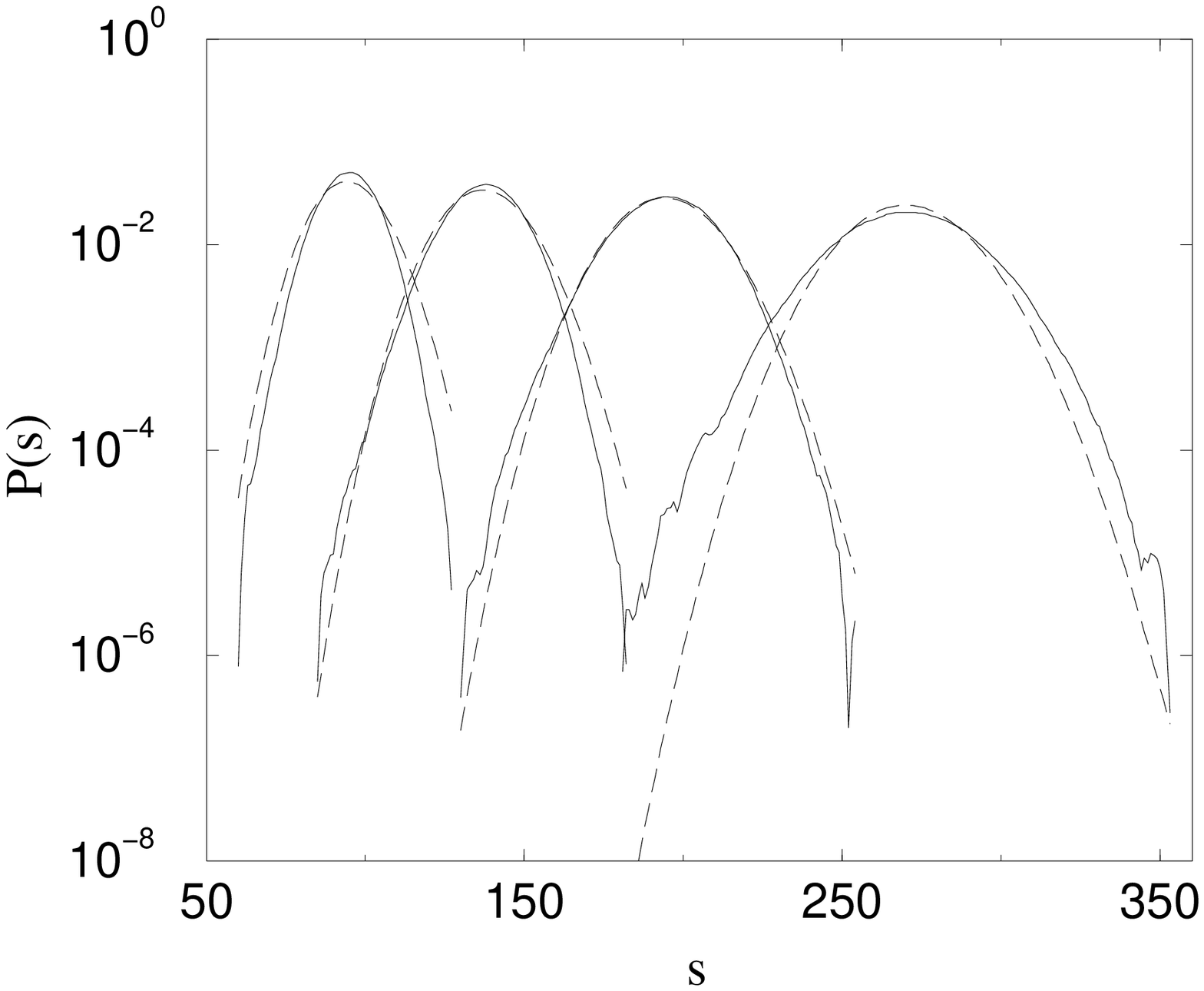,width=8cm}
}
\caption{(a) Ratio between the variance and the mean of the distribution of
the number of species. (b) Comparison of four stationary $P(s)$ (solid
lines) with Poissonian distributions with the same average value (dashed 
lines). From left to right resources and immigration rate increase with the 
other parameters fixed. Model (B) of the continuous dynamics has been used.}
\label{fig:var}
\end{figure}

We also measured the whole stationary distribution of the number of species
$s$, $P(s)$. We plot in Fig.\ \ref{fig:var}a
the ratio between the variance $V_s=\la s^2\ra-\la s\ra^2$ and the mean
$S=\la s\ra$
as a function of the mean number of species $S$.
The measure is much noisier than that of the
average value, and it is strongly affected from possible sampling errors.
The variance is typically
lower than the mean for small $\la s\ra$, but at some
point increases faster than the mean, becoming larger than it at
large $\la s\ra$. Since for a Poissonian
distribution the variance and the mean are equal, the distribution $P(s)$
is narrower than a Poissonian one for small $\la s\ra$ and broader than it
for large $\la s\ra$ (see Fig.\ \ref{fig:var}b, where for each curve a
Poissonian distribution with the same mean value has been represented as a
dashed line). Notice however that the comparison with a Poissonian
distribution is very good for some range of parameters.

\subsection{Species area relationship}
\label{area}

In order to investigate the dependence between biodiversity and area,
we have to fix the relationship between area and immigration rate $I$.
Usually, a positive correlation is expected even though
the actual dependence may vary with the species considered. We restrict
our study to the assumption that the immigration rate is proportional to the
linear  size of the island (MacArthur \& Wilson, 1963):

\be I=1/T_\mig=I_0+k A^{1/2}\: .\label{immi-scale}\ee

We also include a constant $I_0$ to take into account that, for
islands in an archipelago, the immigration rate depends much more on
the geometry of the archipelago and on its distance from the mainland
than on the value of the area. For isolated islands and whole archipelagos,
on the other hand, there is usually a unique source of immigrants from
the mainland, and the effect of area on immigration
is expected to be important. In view of this situation,
in the first part of the discussion the constant $I_0$ will be neglected.
The parameter $k$ is related to the distance from the mainland.

In the framework of the IBM, the other parameter influenced by area
is the number of patches $N_h$, which is taken to be proportional to area:
$N_h=A$, with an appropriate choice of units. Thus we simulated systems
with different values of $N_h$, varying the immigration rate as above.
Our main result is that we always obtain a power-law dependence of the
number of species with the area, with typical values of $z$ in the
interval $0.6-0.8$, as it is observed for the case of archipelagos, to which
our immigration model should apply.
Taking into account additional sources of immigrations, like
closeby islands, is expected to reduce
the dependence of the immigration rate on area, and thus to cause a
decrease of the effective $z$ values, making them more similar to the values
observed in groups of neighboring islands.

\begin{figure}
\centerline{
\psfig{file=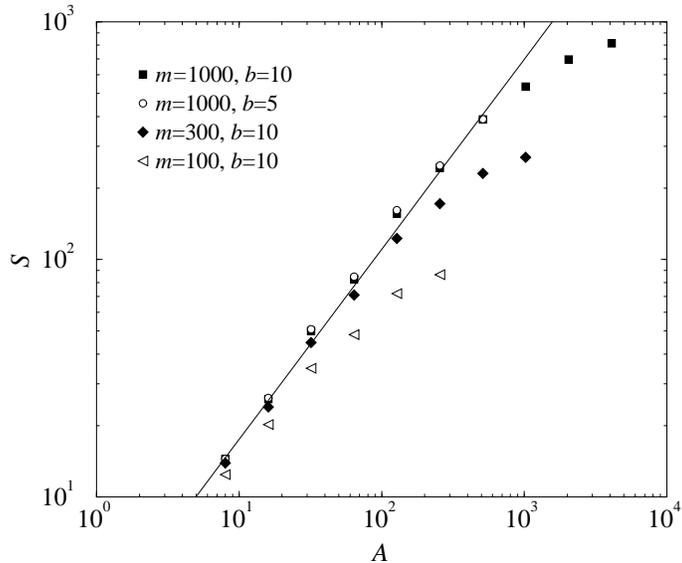,width=9cm,angle=270}}
\caption{Average number of species as a function of area in the IBM model,
for an immigration rate proportional to the square root of $A$. The
three curves correspond to different sizes of the species pool $M$, as 
shown in the legend. The remaining parameters are equal for the three curves
and are $E=150$, $E_{rep}=200$, $d=1$, $\delta=20$, $p_d=0.002$. The straight
line is given by $z\sim 0.75$.
This results
are obtained with the cascade model. Similar curves are obtained for a random 
matrix, with values of $S$ slightly below these, on the average.}
\label{fig:IBM4}
\end{figure}

For the IBM, we represent in Fig.\ \ref{fig:IBM4} the number of species as
a function of area, with immigration rate $I \equiv k \sqrt{A}$ and
$N_h\equiv A$, for different values of the continental pool $M$.
It can be seen that the species area relationship bends for large areas,
apparently tending to an asymptote. Increasing the species pool $M$
increases the asymptote, but leaves the value of the effective exponent
unchanged.
Parameter values are $E=150$, $E_{rep}=200$, $d=1$, $\delta=20$,
$p_d=0.002$ and $k=0.353$.

\vspace{.5cm}

We now come to the species-area relationship in the continuous formulation
of the ecological dynamics.
We should first discuss how the parameters of the ecological
equations, Eqs.\ (\ref{ParaA}) and (\ref{ParaB}), depend on area.
The variables $N_i$ of ecological equations have the meaning of
spatial densities of individuals. Thus the equations Eq.\ (\ref{equ})
are invariant with respect to changes of area. There is however
another important determinant of the ecological dynamics: The threshold
$N_c$ below which extinctions happen. Two different cases have to be
considered:

\begin{enumerate}

\item $N_c$ independent of area, in other words there is a critical
density below which the species go extinct, as it happens for the Allee's
effect (Allee {\it et al.}, 1949). Such a
situation is expected, for instance, if the individual of the
species are uniformly dispersed in the area $A$ so that, below the critical
density $N_c$, they can not find mating partners.

In this case, the ecological dynamics is
invariant under changes of the area and, in particular, the
extinction rate does not depend on $A$. Assuming that the
immigration rate increases with area as in Eq.\ (\ref{immi-scale}),
and that the area is much larger than $(I_0/k)^2$, we find

\be 
S\propto A^z\:, \:\:\:\:\: z=a/2\:,
\label{nonIA}
\ee
where $a$ is the exponent in Eq.\ (\ref{S-I0}).

\item Extinctions depend on the absolute
number of individuals, $N_i A$ and the extinction threshold is thus

\be N_c\propto 1/A\, .\label{scale}
\ee

\begin{enumerate}
\item First we consider this case together with $I=I_0$ independent of area
(corresponding to $k=0$ in Eq.\ (\ref{immi-scale})). From equation
(\ref{S-log}), we obtain then that the number of species increases
logarithmically with area:

\be S\approx A_1+A_2 \log(A)\, .
\ee
This relation is indeed observed for birds in the central islands of the
Solomon archipelago, which are all very close to at least one other
large central islands (Diamond \& Mayr, 1976). For such islands, the
immigration rate can be expected to be rather independent of area.

\item
Extinctions depend on the absolute number of individuals and the
immigration rate increases with island size $(A\gg (I_0/k)^2)$,

\be N_c\propto 1/A\, ,\:\:\:\:\:\: I=kA^{1/2}\, .\label{case3}
\ee
In this case, we can not rely on previous results, and we have to
perform new simulations, scaling the parameters as in Eq.\ (\ref{case3}).
\end{enumerate}
\end{enumerate}

We note that, in both cases 2(a) and 2(b), model (A) can become problematic
at large area.
In fact, as area increases the coupling constant $\g_\max N_c/\a\propto 1/A$
becomes smaller, and the system becomes dominated by dissipative effects.
The result is that, unless $\alpha=0$, Eq.\ (\ref{Ediacara}) in the Appendix
would not be satisfied as area increases, and the system would meet a
``garden regime", in which only basal species survive (see Appendix). 
It is not surprising that Lotka-Volterra equations go in trouble for very
low densities: in fact, they are analogous to equations of chemical kynetics,
and when the density of the species involved becomes too small, the assumptions
on which they are founded break down.
To avoid such a problem, we used the ratio dependent model (B) together
with Eq.\ (\ref{case3}). Nevertheless, our numerical study shows that also
model (A) provides comparable results in a suitable range of parameters.

\begin{figure}
\centerline{
\psfig{file=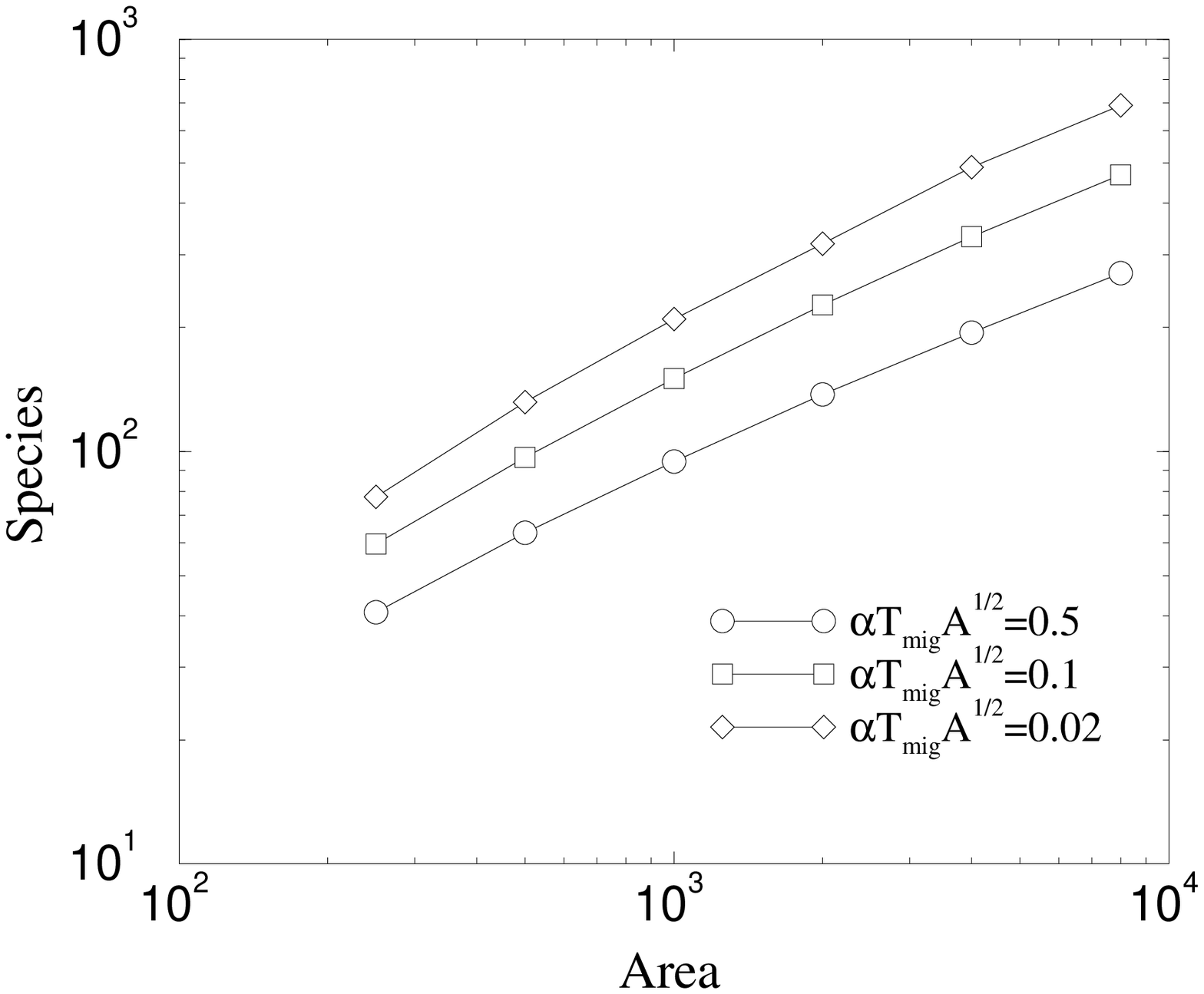,width=8cm}
\psfig{file=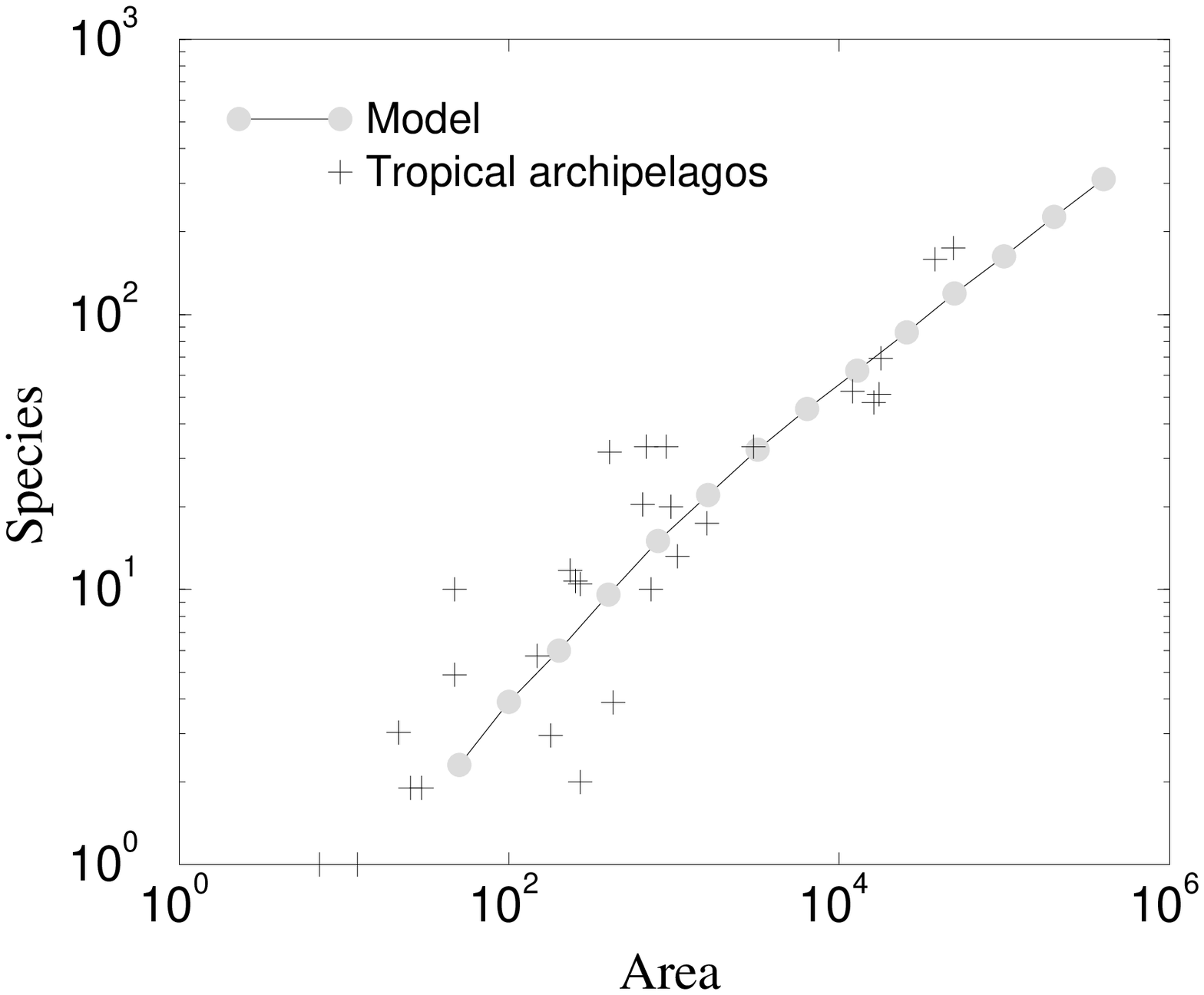,width=8cm}}
\caption{Number of species as a function of area, in the hypothesis
$R\propto A$, $T_\mig= kA^{-1/2}$. (a) Curves for different values of
$k$. (b) A curve for $k=0.02$ compared to biodiversity data of Pacific
archipelagos (Adler, 1992; Rosenzweig, 1995). The frequency dependent 
continuous equations, Eq.\ (\ref{denom}), have been used. Empty symbols 
refer to $\ell_\max=4$, filled symbols to $\ell_\max=8$.}
\label{fig:area}
\end{figure}

Results are plotted in Fig.\ \ref{fig:area}, and yield an approximately
power law species area relationship, for large enough values of the
parameter $\a k$. In the log-log plot, the curves $S(A)$ show a negative
curvature, that can be eliminated introducing a new parameter $A_0$,

\be S\approx c(k)\l(A-A_0(k)\r)^{z(k)}\: .\label{sar2}
\ee
Both the effective exponent $z(k)$ and the limit area $A_0(k)$ increase
slowly with the immigration parameter $k$. For the curves in Fig.\ 
\ref{fig:area}, the exponent $z(k)$ ranges from $z=0.49$ at $k=2$ to 
$z=0.56$ at $k=50$.

We notice however that the scaling form Eq.\ (\ref{sar2}) is only
approximate, that a scaling of the form $A^z\log(A)$ would probably be
more adequate, and that the immigration rate is a better scaling
variable than area. We extract an exponent from
our approximately power law species area relationship only for the sake
of comparison with field data.

Our model of immigrations applies to whole archipelagos or to
isolated islands, because we consider a unique source of immigrations.
It is remarkable that the exponent $z$ observed for a set of Pacific
archipelagos has the value $z=0.54$ (Adler, 1992;
Rosenzweig, 1995), in very good agreement with the results of our
simulations. Real data are shown for comparison in Fig.\ \ref{fig:area}b.

It is striking that models at different description levels, as the IBM
and the continuous model (B), yield very similar species area relationships.
We compared other statistical features of the individual based and the
continuous models, finding that they are qualitatively very similar
(this holds for the Lotka-Volterra model as long as the garden regime
and the opposite low dissipation regime are avoided).
In the following we will present a complete analysis of the stationary state
for data obtained from the IBM and from model (B).

\begin{figure}
\centerline{
\psfig{file=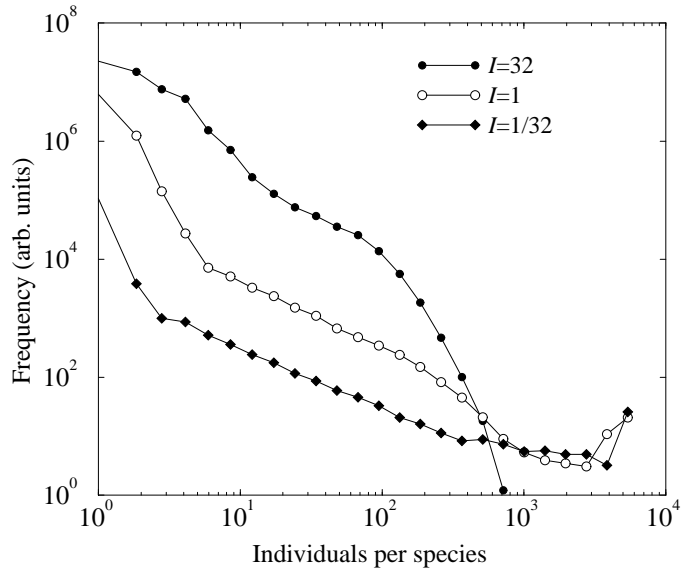,width=9cm,angle=270}}
\caption{Distribution of the number of individuals per species in the IBM
model for different values of the immigration rate. The remaining values are
as in Fig.\ \ref{fig:IBM2}. As
can be seen, the abundances of species are power-law distributed with an
exponent close to unity in all cases. The increase in the immigration rate 
moves the exponential cut-off at large sizes towards the left.}
\label{fig:IBM3}
\end{figure}

\subsection{Distribution of species abundances}

We measured distributions of species abundances, defined as the
probability density $p (N)$ of species with $N$ individuals
(or total biomass equal to $N$ in the case of the continuous model),
both for the IBM and for the continuous model. We observe a good qualitative
agreement of the results in both approaches.

In the framework of the IBM, we measured the distribution of species
abundances for three values of the immigration rate corresponding to the
three regimes in Fig.\ \ref{fig:IBM2} (slow, intermediate and fast driving).
Fig.\ \ref{fig:IBM3} represents the frequency $p(N)$ with which
species formed by $N$ individuals were recorded.
Different curves refer to different immigration rates (other parameters as in 
Fig.\ \ref{fig:IBM2}). All curves show an initial 
fast decaying part corresponding to species that go extinct almost 
immediately after arriving to the island. Since these species do not find  
preys to feed on, their initial energy decays exponentially and they die 
out of starvation. The relevant part of these distributions results from
species which play a role in the ecological network. This part shows a
power-law decay of the form

\begin{equation}
p(N) \propto N^{-\xi} \; ,
\end{equation}
with $\xi \simeq 1$.
Finally, the external
resources set the value of $N$ at
which an exponential cut-off appears. Our results are in good agreement with
field measures of diversity, many of which also return a power-law distribution
of species abundances with an exponent in the range $1-1.25$ (Pielou, 1969;
Sol\'e {\it et al.}, 2000).

\begin{figure}
\centerline{
\psfig{file=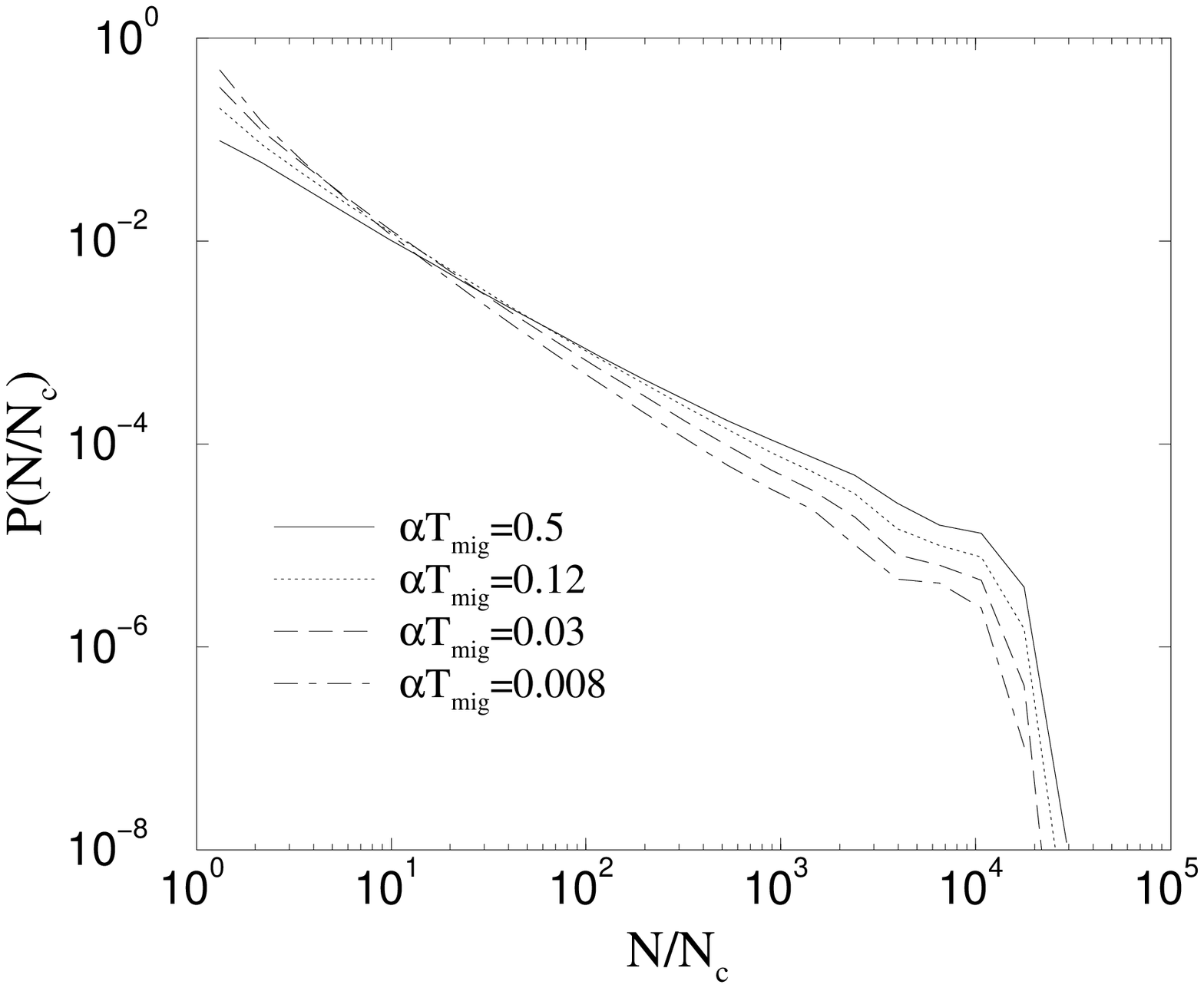,width=8cm}
\psfig{file=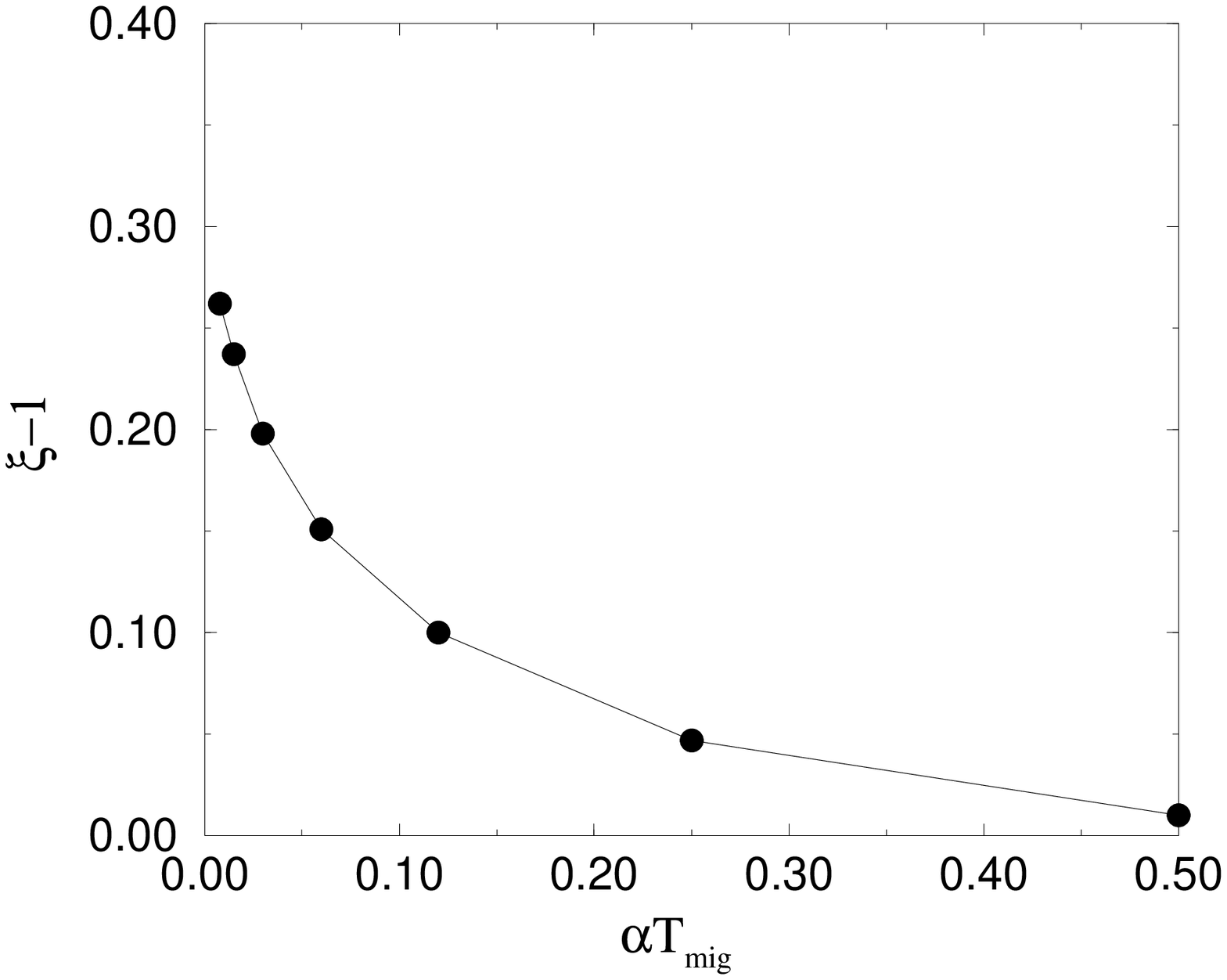,width=8cm}
}
\caption{(a) Distribution of the biomass per species in the continuous
model (B), for different values of the immigration rate.
Other parameters: $N'_c=10^5$, $\b'=1$, $c'_\max=2$, $\ell_\max=4$.
(b) The exponent $\xi$ of the power-law part of the distribution
minus one as a function of the immigration rate. Same parameters as in
the previous part.}
\label{fig:size}
\end{figure}

The same results are obtained in the framework of the continuous model.
In this case, however, we observe that the exponent $\xi$ increases slowly with
immigration rate, tending to $\xi\approx 1$ in the limit $I\to 0$.
The maximum value that we found in our simulations is $\xi\approx 1.25$, still
compatible with observational data. A sample of results is reported in
Fig.\ \ref{fig:size}a. In Fig.\ \ref{fig:size}b the decrease of the
exponent $\xi$ with the immigration rate is shown.

\subsection{Lifetime distribution}

\begin{figure}
\centerline{
\psfig{file=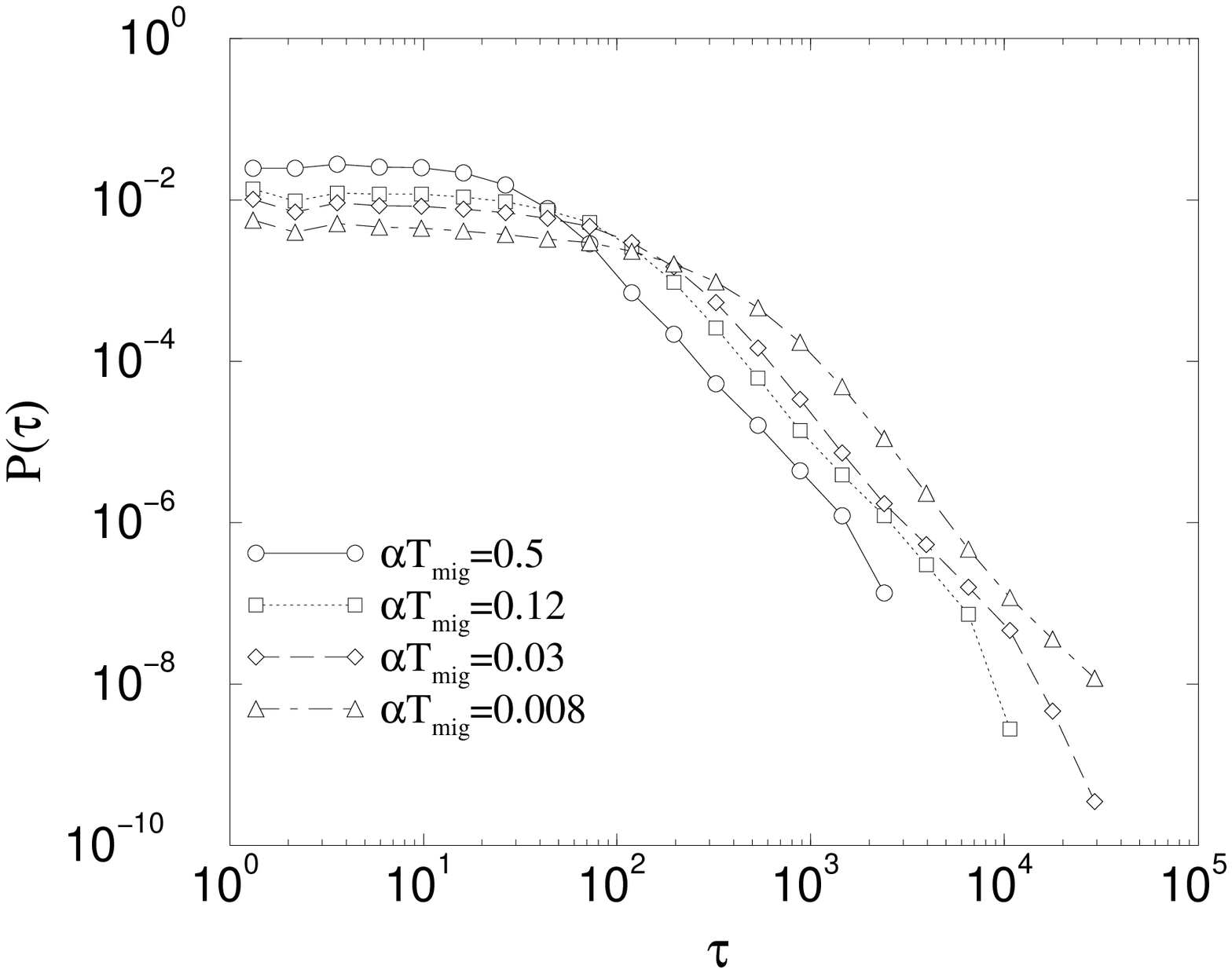,width=8cm}
\psfig{file=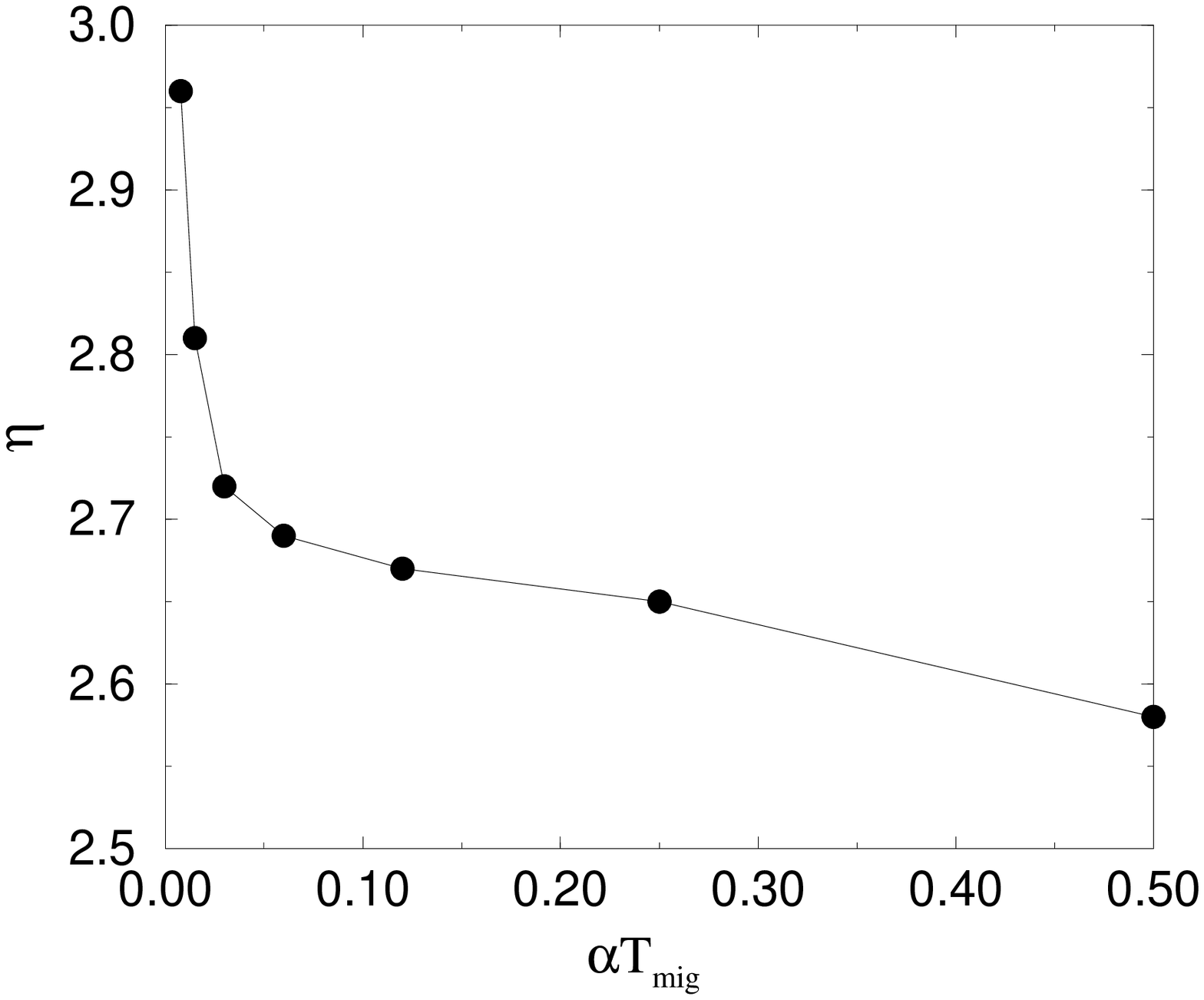,width=8cm}
}
\caption{(a) Distribution of the lifetimes in the continuous
model (B), for different values of the immigration rate.
Other parameters: $N'_c=10^5$, $\b'=1$, $c'_\max=2$, $\ell_\max=4$.
(b) The exponent $\eta$ of the power-law part of the distribution
as a function of the immigration rate. Same parameters as in
the previous part.}
\label{fig:tau}
\end{figure}

The distribution of lifetimes of species is shown in Fig.\ \ref{fig:tau} in
a log-log plot, for several values of the immigration rate and of other
parameters. After an initial part where the distribution is almost uniform,
corresponding to species with very short permanence time, we observe
an approximately power law decay of the probability density
for a range of at least one and half decade

\be p(\tau)\approx \tau^{-\eta} \label{p_tau}\: ,
\ee
when using different parameter values,
we found values of the effective exponent $\eta$ between
2.1 and 2.8.

The average lifetime $\la \tau\ra$ in the equilibrium state is related
to biodiversity through the relation

\be \la \tau\ra =T_\mig \la S\ra \propto T_\mig^{b/(1+b)},
\ee
which follows from its definition and from the properties of the
stationary state. Here $T_\mig$ represents the average
time between succesful immigrations which contribute to the island
biodiversity. Thus the average lifetime decreases with the immigration rate
$1/T_\mig$ and, consistently, the value of the
exponent increases, as it is shown in Fig.\ \ref{fig:tau}b.

Our results compare qualitatively well with observed patterns
(Keitt \& Marquet, 1996; Keitt \& Stanley, 1998).
It was in fact observed that the time of permanence of birds in local
patches follow a distribution approximately of the form
(\ref{p_tau}) with effective exponent $\eta=1.6$, indeed smaller than the
typical values found in our simulations.
A result which compares better to this last value has been found, using a 
model without explicit ecological dynamics, in (Sol\'e {\it et al.}, 2000).

\subsection{Network organization}

The structure of the ecological network does change, even if very slowly,
with changing immigration rate. We have examined in particular the
number of trophic levels, the number of links per species and the total
biomass.

\begin{figure}
\centerline{
\psfig{file=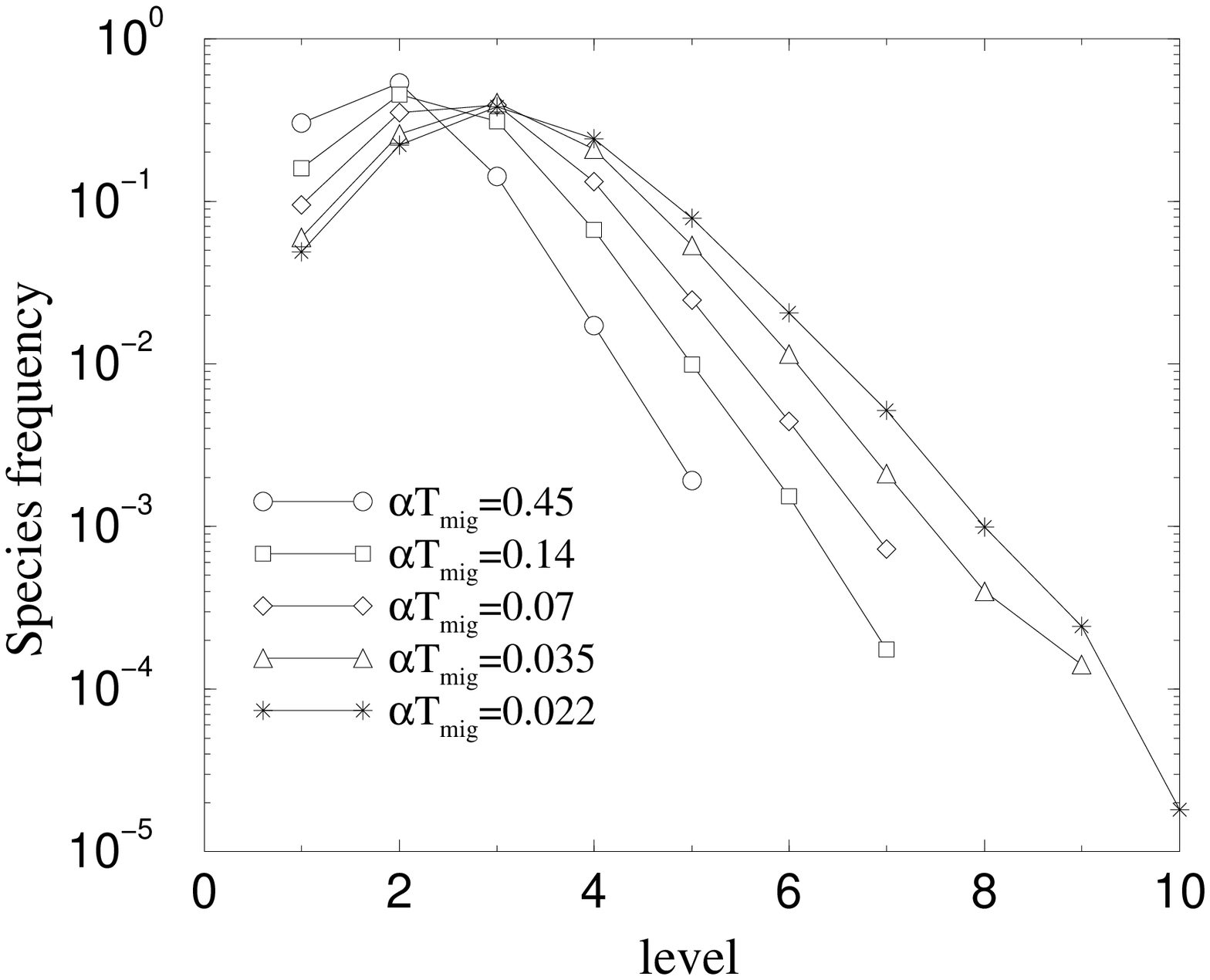,width=8cm}
\psfig{file=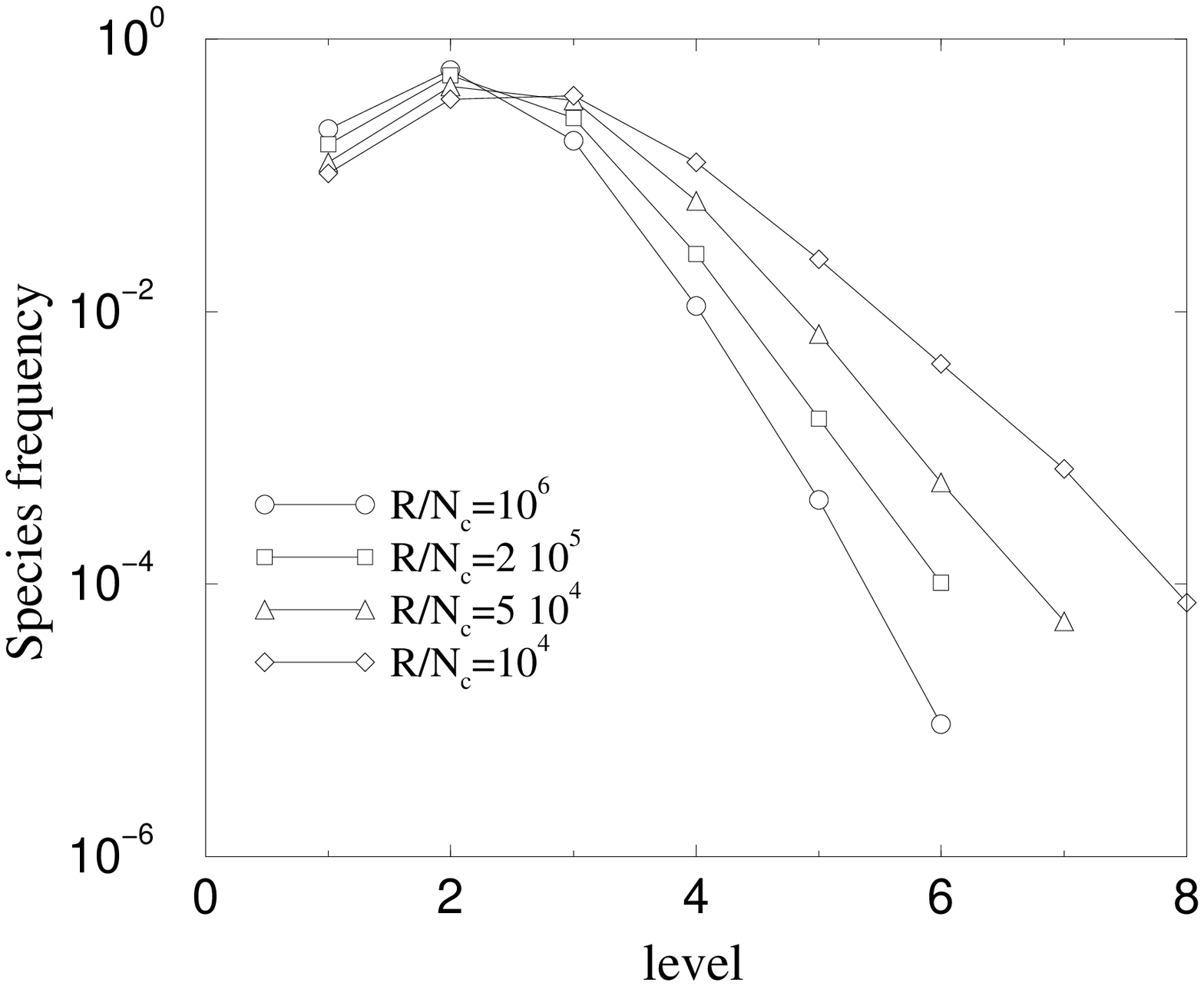,width=8cm}
}
\caption{(a) Frequency of species at trophic level $l$ in all the course
of the simulation in the continuous model (B), for
different values of the immigration rate. Food webs become longer at
increasing immigration rate. (b) Same, for different values of the resources.}
\label{fig:level}
\end{figure}

We define the trophic level of a species as the minimal number of links
connecting it to resources.
In all our simulations the number of trophic levels varies between
four and ten. It shows a tendency to increase with immigration rate,
as it is illustrated in Fig.\ \ref{fig:level}.

\begin{figure}
\centerline{
\psfig{file=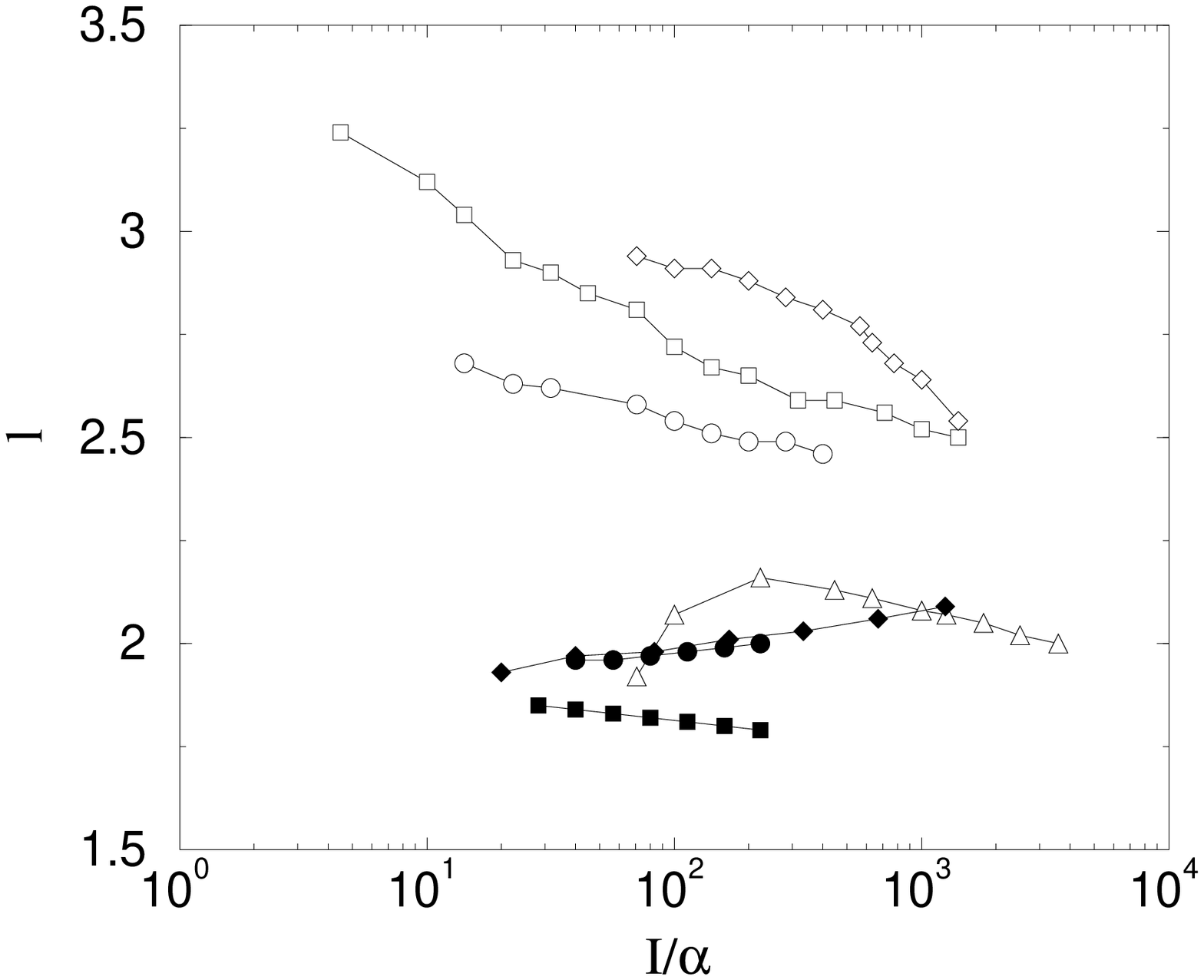,width=8cm}
\psfig{file=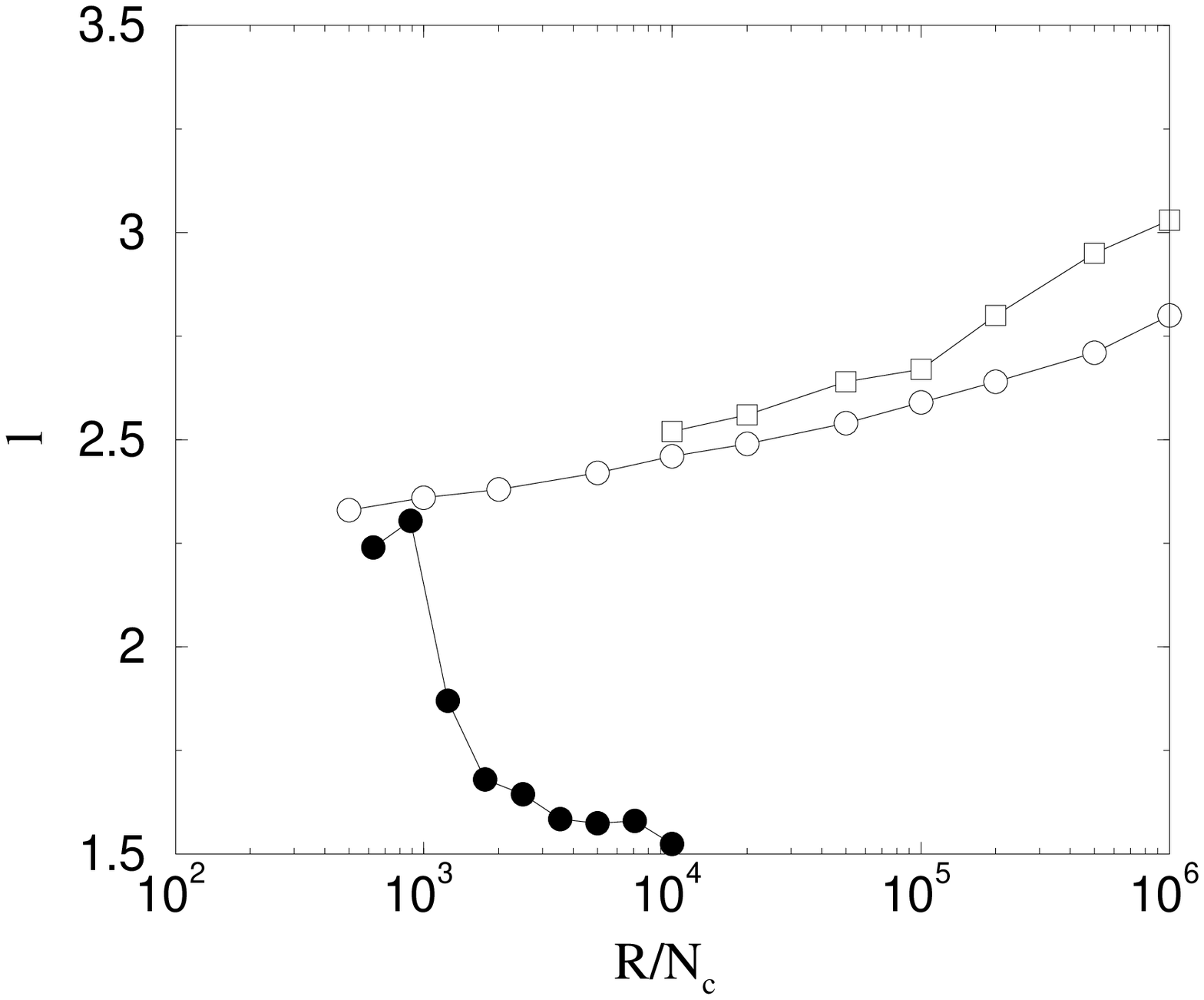,width=8cm}
}
\caption{(a) Number of links per species as a function of the immigration
rate for the continuous model (B). Empty points refer to $\ell_\max=8$,
filled points to $\ell_\max=4$.
(b) Same, as a function of the resources $R$.}
\label{fig:link}
\end{figure}

The average number of links per species, counted as average number of preys, 
is shown in Fig.\ \ref{fig:link} as a
function of the immigration rate. It changes very slowly (logarithmically)
and, in some cases, in a non monotonic way (for most curves we only observe
either the increasing or the decreasing part). A similar pattern is observed
as a function of the resources $R$ (see Fig.\ \ref{fig:link}b).
Thus, as a function of the number of species, the number of links per species
behave non monotonically. It also depends weakly
on the maximum number of links allowed when the new species is added to the
ecosystem, $\ell_\max$.

\begin{figure}
\centerline{
\psfig{file=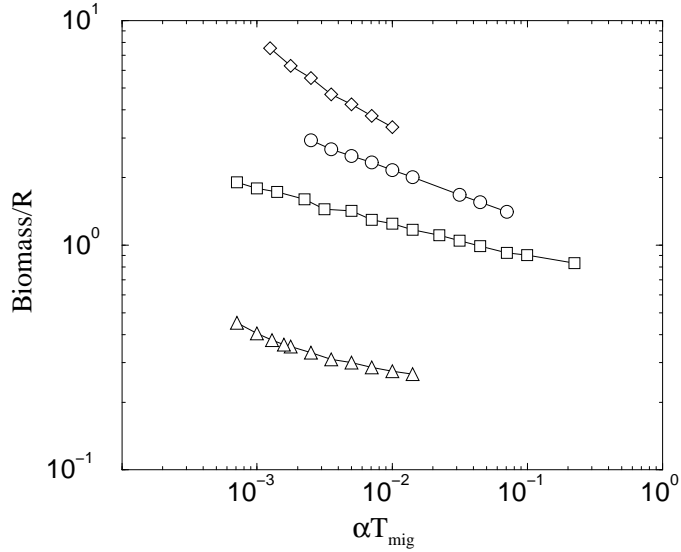,width=9cm}
}
\caption{Biomass as a function of the inverse of the immigration rate,
$T_\mig$, in the continuous model (B).}
\label{fig:bio}
\end{figure}

The total biomass also increases approximately as a power law of the
immigration rate, as shown in Fig.\ \ref{fig:bio}. The exponent ranges from
0.15 to 0.58.

\section{Relationship to MacArthur \& Wilson's theory}\label{phenom}

We have already seen in the previous section that quantitative biodiversity 
patterns can be derived from a balance between external driving through 
immigrations and the intrinsic population dynamics of the ecosystem.
Here we want to relate these results to existing phenomenological
approaches, in particular  MacArthur and Wilson's (1963, 1967) theory
of island biogeography. 

In an ensemble average (or, equivalently, in an average over long times),
the response of the system to a constant immigration flux $I$ reduces 
to a stationary extinction $E$. This can be expressed as a function of the
avergage number of species $S$ in the system, $E=E(S)$. 
Of course, this function depends also on the model paramenters and on
qualitative features of the immigration flux.
Since the system
is in a stationary state, immigration and extinction have to compensate 
each other on average:
\beq
I - E(S) = 0 \;.
\label{S_phen}
\eeq
This balance is indeed as postulated by MacArthur and Wilson. 
The immigration flux measures the average number of new species arriving
per unit of time.  The functional form of the `extinction curve' $E(S)$
can now be obtained from the underlying population dynamics. As explained
above, we find 
\beq \label{extinction}
E(S)\, =\, E_0 S^{1+b}\, ,
\eeq
in the scaling regime where the number of species
increases as a power-law of the immigration rate.
The exponent $b$ was called competition exponent and has been introduced
in Eqs.\ (\ref{comp}) and (\ref{comp2}). In fact, from the stationary solution
of Eq.\ (\ref{S_phen}), we recover Eq.\ (\ref{S-I0})
\beq
\label{Sdb}
S\, =\, \l( \frac{I}{E_0}\r)^{\frac{1}{1+b}}\, .
\eeq
This equation allows to derive the exponent $z$ of the species-area
relationship. Indeed, if we assume that

\be I\propto A^{s}\;, \:\:\:\:\:\:\: E_0\propto A^{-\epsilon}\:, \ee
we find
\be 
S\propto A^z\;,\:\:\: \makebox{with} \:\:\:\: z={s+\epsilon\over 1+b}\: .
\ee
For the population models, we assumed $s=1/2$ (the immigration rate is
considered proportional to the linear size of the island), and we
obtained $\epsilon=0$ (in fact, the number of species increases as the
logarithm of $R/N_c$ at fixed $T_\mig$, both in the IBM and in the
continuous model (B)).

We remark that here, as in the explicit population models, we are assuming that
the only source of immigrants is a continent far apart. The exponents that
we find should then be compared to the exponents observed for isolated
islands or archipelagos, while the exponent computed among islands of the
same archipelago is expected to be lower, due to a reduced dependence of
immigration rate on area. Another point to remark is that 
the immigration rate $I$ measures the flux of new species arriving 
on the island. If this flux is assumed to originate from a `continent'
of $M$ species, $I$ can be related to the total immigration flux $I_0$
by correcting for the immigrations already present on the island. The 
simplest ansatz is  $I(S)=I_0 (1-S/M)$ (MacArthur and Wilson, 1963, 1967).
Expressed in terms of $I_0$, the average number of species $S$ reaches
a saturation value of order $M$. (The pool  of 
immigrant species  is indeed finite in our IBM, but infinite in the continuum
models. Hence, this correction becomes important in comparing results
of the different models). 

\section{Summary and Conclusions}

We have presented a study of biodiversity in an insular ecosystem
at the individual and the population level. Our interest has been focused 
on the statistical properties of the dynamical stationary state and on
the scaling relations between the system variables. Instead of describing 
detailed situations in which some particular species and their exact 
interactions with their known preys and predators are included, we let 
ecological networks self-organize through random assemblage of species, 
ecological dynamics and possible extinctions.

Our main result is that, in a broad range of parameters, biodiversity scales
approximately as a power law of the immigration rate. The value of the 
exponent varies slightly when the parameters of the models are changed, but the
qualitative features of the stationary state are quite robust. 

The behavior of biodiversity with immigration rate allows to derive a
species area relationship with a power law shape, if we assume that
the immigration rate is proportional to the linear size of the island.
Such a model of immigration considers as unique source of diversity a flux
of species from a continent far apart, thus it should be compared to
observations relative to isolated islands or to whole archipelagos.
The agreement is in this case rather good: the observed value of the
effective exponent $z$ on archipelagos is $z=0.54$
(Adler, 1992; Rosenzweig, 1995), while we typically get, with the continuous
model, values between $0.52$ and $0.56$ and, with the individual based
model, values between $0.6$ and $0.8$.
Thus, the comparison of our two description levels points out to 
species-area law of the type (\ref{SAz}) as a generic feature of a broad set
of ecological models with random interactions.

Our models reproduce qualitatively other features observed in real ecosystems.
We observe a power law distribution of population abundances, {\it i.e.} the 
number $p(N)$ of species with $N$ individuals approximately decreases as 
$p(N) \propto N^{-\xi}$. This is expected to be a general consequence of the 
multiplicative nature of population dynamics equations. The exponent $\xi$ 
found in our simulations is close to unity, in favourable comparison with 
field data, and increases with the immigration rate.

We also observe a broad distribution of the time of permanence of species
in the system $\tau$, as it has been observed in the field (Keitt \& Marquet,
1996; Keitt \& Stanley, 1998) and in a related model (Sol\'e {\it et al.}, 
2000). The average permanence time is proportional to the number
of species and inversely proportional to the immigration rate,
$\la \tau\ra= \la S\ra T_\mig$, so that it decreases with the immigration rate.
The fact that it is observed, both in field studies and in models, that its 
distribution is broad, could help to reconcile the apparent dichotomy between 
fugitive species and permanent species: These groups of species could
correspond to the two extreme cases of a unique distribution of permanence
times (Schoener \& Spiller, 1987). The approximate power law shape
of the distribution of times of permanence in the island is
reminiscent of the analogous distribution of the lifetime of genera in the
fossil record, which is approximately given by $P(\tau)\propto \tau^{-\eta}$,
with $\eta \approx 2$, close to what is observed in our model for very
small immigration rates and also close to ecological observations.
It is tempting to speculate that this similarity points out at similar
mechanisms acting on the time scales of ecosystem dynamics as on the
timescales of macroevolution.

The number of trophic levels in the food web is also strongly influenced
by the immigration rate. We typically find from four to ten
trophic levels, depending on parameters, and with a tendency for the
number of levels to increase with immigration rate. Hints to the
correlation between immigration rate and number of levels can be found
in the fact that the length of food chains appears to be positively correlated
to habitat area, although the data are quite poor
(Schoener, 1989; Spencer, 1997). Our results suggest that one of the factors
limiting the length of food chains is the immigration or speciation rate.
Notice that in our model no other limitations to the length of food chains
exist: energetic considerations would limit the number of levels to a value
$\log(R/N_c)$, much larger than the one observed.

An important result of our study
is that the observed statistical patterns are rather robust with respect to
changes in the dynamical rules of the model.
One example is the representation of space in the IBM model. Although
one could think that in this case explicit space is needed, we modelled
space only in an effective way, increasing the immigration rate and the 
resources with the area. We believe that this effective approach captures
the main features of the behavior of biodiversity with area, even if
important issues, like for instance the presence of many different
habitats, are not represented in the model.

As Pimm
poses it, {\em (...) it is pointless to try to justify models' equations
biologically -- their assumptions are almost bound to be wrong. (...) The 
concern should not be whether the assumptions are wrong (they are!), but 
whether it matters that they are wrong.} (Pimm, 1991).
It seems that the statistical laws and the scaling relationships that
we observed
are generic properties of complex ecosystems, that is an unavoidable results
of a minimal set of rules governing population dynamics and
immigrations. Thus the strategy
is to look for the simplest set of rules which appear sensible
and which allow to derive the observed statistical patterns of biodiversity.

\section*{Acknowledgments}
Discussions with David Alonso, Lloyd Demetrius, Barbara Drossel,
Lorenz Fahse and Martin Rost are gratefully acknowledged.

\section*{Appendix. The garden regime}\label{sec:garden}

\begin{figure}
\centerline{
\psfig{file=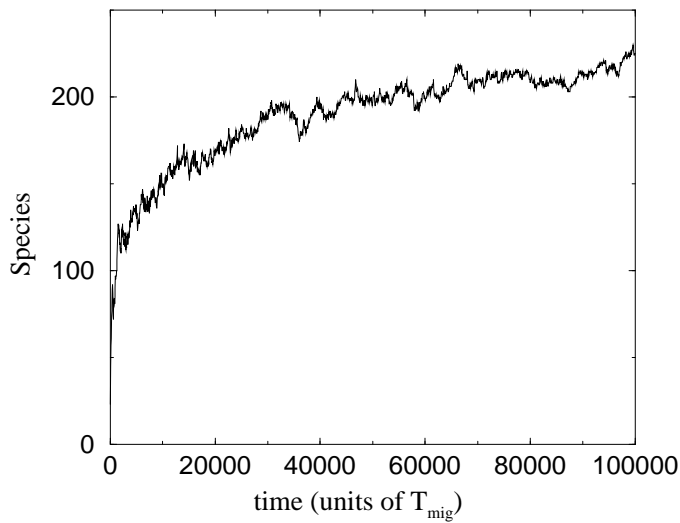,width=9cm}}
\caption{Evolution of the number of species in the garden regime. Notice
that, although the rate of increase gets slower, the number of species
does not reach any stationary value yet. Continuous equations with
constant $\g_{ij}$'s have been used. Parameters: $\g N_c/\a=0.1$,
$\b'=10$, $N'_c=1000$, $\ell_\max=4$.}
\label{fig:garden}
\end{figure}

In the case of model (A) of continuous dynamics,
we observed that only basal species
could survive on the long run for parameter values in a certain range.
In this case, observing the system on a
very long time scale, we did not see any stationary state
(although a stationary state was reached for a much smaller system), but we
saw a number of species steadily and slowly increasing in time
(see Fig.\ \ref{fig:garden}).

We call such a regime the {\it garden regime}, since predators are
absent. Its statistical properties are peculiar: the distribution of
biomasses is narrow and peaked at very low values, and the distribution of
lifetimes is bimodal, with a high peak for very short lifetimes corresponding
to non-basal species, and a shorter
one for large lifetimes, corresponding to basal species. (Even if at any moment
there are many more basal species than predators, the number of predators
passing through the system and almost immediately going extinct is much
larger than the corresponding number of basal species.) Finally, the
distribution of the transfer rate $\g_{0i}$ from the external resources
to species $i$ is strongly peaked close to the maximum allowed value
$\g_\max$. All these features can be easily rationalized as follows.

Let us consider a situation where $S$ basal species coexist feeding on the
abiotic resources $N_0$. Our calculations (unpublished) show that
at the static fixed point of the corresponding
ecological equations, Eqs.\ (\ref{equ}) and (\ref{Res2}), all biomasses are
positive if and only if all differences in the coefficients $\g_{0i}$
are smaller than
$B/S$, where $B=\g_0\b R/\a$. Thus, as $S$ increases, the coupling
constants between the basal species and the environment deviate from
the initial distribution and become more and more similar.
Notice that, if $\b=0$, the coefficients $\g_{0i}$'s should be exactly
equal to guarantee coexistence (this expresses in this context the
``principle of competitive exclusion'').
This conclusion does not vary qualitatively if one considers
Eq.\ (\ref{denom}) instead of the model with constant
coefficients $\g_{ij}$. Thus, in a
system without predators and high immigration of basal species,
we would expect to find many basal species with
very similar biomasses, all of order $R/S$ ($S$ should be smaller than
$R/N_c$) and coefficients $\g_{0i}$ very close one to each other in value.

Let us now consider the arrival of a predator to such a system,
considering first Lotka-Volterra equations (model (A)).
The rate of growth of the predator cannot be larger than $r_\max$ given by

\be r_\max =\ell_\max \g_\max {R\over S}-\a-\b N_c.
\ee
Since $R/S$ is larger than $N_c$, we find that, for large $S$,
a non-basal species can colonize only if
\be 
\g_\max\geq {1\over \ell_\max}\l({\a\over N_c}+\b \r)\, .
\label{Ediacara}
\ee
Every time we observed in our simulations ecosystems mainly composed of
basal species and whose biodiversity has a time behavior similar to that
in Fig.\ \ref{fig:garden}, the above condition was not fulfilled.

In model (B), Eq.\ (\ref{denom}), $r_\max$ equals $\ell_\max \a' c_\max-\a
-\b N_c$. This quantity must be larger than zero, otherwise no
species would survive. Thus the ``garden regime" that we described in this
appendix can not be found in simulations of ratio dependent functional
responses, unless we decide to use different parameters for basal and non-basal
species.

\newpage

\begin{table}
\label{tab1}
\begin{center}
\begin{tabular}{lll} 
\bf Parameter & \bf Meaning & \bf Value-Range \\ 
 \hline
$p_d$ & Death probability & 0.001-0.02 \\ \hline 
$d$ & Dissipation rate & 1-10 \\ \hline 
$N_h$ & Max. Ind. in basal species ($\simeq$ island area) & 100-10000 \\ 
 \hline
$b$ & Basal growth & 2-25 \\\hline 
$C(m_i,m_j)$ & Energy obtained by $i$ when eating $j$ & max. 
$E \in [20,200]$ \\ \hline 
$E_{rep}$ & Reproduction energy & 100-300 ($E_{rep} \approx 
2 E$) \\ \hline
$\delta$ & Energy when born & 5-$E/2$ \\ \hline 
$\ell$ & Number of preys per predator & 3-4 \\ \hline
$M_h$ & Max. number of basal species & 1-100 or CM\\ \hline
$M_a$ & Max. number of animal species & 1-1000 \\ \hline
$I$ & Immigration rate & $10^{-3} - 10^3$ \\ 
\end{tabular}
\end{center}
\caption{Parameters in the IB model and values used.}
\end{table}


\end{document}